\documentclass{aa}  

\usepackage{graphicx}
\usepackage{caption}
\usepackage{subcaption}
\usepackage{comment}
\usepackage{txfonts}
\usepackage[bookmarks=false,colorlinks=true,linkcolor=black,citecolor=cyan,filecolor=black,urlcolor=cyan]{hyperref}
\usepackage{xcolor}

\begin{document}

   \title{The hot circumgalactic medium in the eROSITA All-Sky Survey}
   \subtitle{II. Scaling relations between X-ray luminosity and galaxies' mass}

   \author{Yi Zhang \inst{1}\fnmsep\thanks{yizhang@mpe.mpg.de} \and
    Johan Comparat\inst{1} \and Gabriele Ponti\inst{2,1} \and Andrea Merloni\inst{1}
    \and Kirpal Nandra\inst{1}
    \and Frank Haberl\inst{1}
    \and Nhut Truong \inst{3,4}
    \and Annalisa Pillepich\inst{5}
    \and Nicola Locatelli\inst{2}
    \and Xiaoyuan Zhang\inst{1}
    \and Jeremy Sanders\inst{1}
    \and Xueying Zheng\inst{1}
    \and Ang Liu\inst{1}
    \and Paola Popesso\inst{6}
    \and Teng Liu\inst{7,8}
    \and Peter Predehl\inst{1}
    \and Mara Salvato\inst{1}
    \and Soumya Shreeram\inst{1}
    \and Michael C. H. Yeung\inst{1}
          }
          
   \institute{Max-Planck-Institut für extraterrestrische Physik (MPE), Gießenbachstraße 1, D-85748 Garching bei München, Germany
   \and
    INAF-Osservatorio Astronomico di Brera, Via E. Bianchi 46, I-23807 Merate (LC), Italy 
    \and
    NASA Goddard Space Flight Center, Greenbelt, MD 20771, USA 
    \and
     Center for Space Sciences and Technology, University of Maryland, 1000 Hilltop Circle, Baltimore, MD 21250, USA 
    \and
    Max-Planck-Institut für Astronomie, Königstuhl 17, 69117 Heidelberg, Germany 
    \and 
    European Southern Observatory, Karl Schwarzschildstrasse 2, D-85748 Garching bei München, Germany 
    \and
    Department of Astronomy, University of Science and Technology of China, Hefei 230026, China 
    \and 
    School of Astronomy and Space Science, University of Science and Technology of China, Hefei 230026, China
             }
   \date{Received ; accepted }

 
  \abstract
   {}
   {Understanding how the properties of galaxies relate to the properties of the hot circum-galactic medium (CGM) around them can constrain galaxy evolution models. We aim to measure the scaling relations between the X-ray luminosity of the hot CGM and the fundamental properties (stellar mass and halo mass) of a galaxy.}
   {We measured the X-ray luminosity of the hot CGM based on the surface brightness profiles of central galaxy samples measured from Spectrum Roentgen Gamma (SRG)/eROSITA all-sky survey data. We related the X-ray luminosity to the galaxies' stellar and halo mass, and we compared the observed relations to the self-similar model and intrinsic (i.e., not forward-modeled) output of the IllustrisTNG, EAGLE, and SIMBA simulations.}
   {The average hot CGM X-ray luminosity ($L_{\rm X,CGM}$) correlates with the galaxy's stellar mass ($M_*$). It increases from $(1.6 \pm 2.1)\times10^{39} \rm erg\,s^{-1}$ to $(3.4 \pm 0.3)\times10^{41} \rm erg\,s^{-1}$, when $\log(M_*)$ increases from 10.0 to 11.5. A power law describes the correlation as $\log(L_{\rm X,CGM})= (2.4\pm 0.1)\times \log(M_*)+(14.6\pm1.5)$. 
   The hot CGM X-ray luminosity as a function of halo mass is measured within $\log(M_{\rm 500c})=11.3-13.7$, extending our knowledge of the scaling relation by more than two orders of magnitude. $L_{\rm X,CGM}$ increases with $M_{\rm 500c}$ from  $(3.0 \pm 1.6)\times10^{39}\ \rm erg\,s^{-1}$ at $\log(M_{\rm 500c})=11.3$ to $(1.3 \pm 0.1)\times10^{42}\ \rm erg\,s^{-1}$ at $\log(M_{\rm 500c})=13.7$. 
   The relation follows a power law of $\log(L_{\rm X,CGM})= (1.32\pm 0.05)\times \log(M_{\rm 500c})+(24.1\pm0.7)$. Our observations highlight the necessity of non-gravitational processes at the galaxy group scale while suggesting these processes are sub-dominant at the galaxy scale. We show that the outputs of current cosmological galaxy simulations generally align with the observational results uncovered here but with possibly important deviations in selected mass ranges.}
   {We explore, at the low mass end, the average scaling relations between the CGM X-ray luminosity and the galaxy's stellar mass or halo mass, which constitutes a new benchmark for galaxy evolution models and feedback processes.}


   \maketitle
%
\section{Introduction}

In the paradigm of hierarchical structure formation driven by gravitation, the primordial density field made of dark matter and baryons undergoes collapse to form virialized objects: halos and galaxies \citep{White1978}. 
Galaxies, halos, and their environments co-evolve, and their properties are correlated \citep{Muldrew2012}. 
For example, \citet{Leauthaud2012b,Leauthaud2012a} and \citet{Coupon2015} observe a tight relation between the stellar mass ($M_*$) of a galaxy and its host dark matter halo mass ($M_{\rm halo}$). Moreover, galaxies follow a well-known main sequence between star formation rate and stellar mass \citep{Whitaker2012,Iyer2018}, and show a correlation between their size and stellar mass \citep{Mowla2019a,Mowla2019b}. 
In this article, we investigate the relation between the total X-ray luminosity of a halo and the stellar or halo mass of the galaxy. 

In the local Universe, for the early-type galaxies (ETGs), the relation between $L_X$ and $M_*$, dynamical mass (baryon plus dark matter), central optical profile of galaxy, and temperature of gas ($T_X$) within $<50$ Mpc has been studied \citep{Kim2013,Kim2015,Goulding2016,Forbes2017}. 
The relations measured support the necessity of the active galactic nucleus (AGN) feedback in massive ETGs and demands of stellar feedback in less massive ETGs \citep{Choi2015}. 
For the late-type galaxies (LTGs), the Chandra survey of nearby ($<30$ Mpc) highly inclined disc galaxies finds correlations between $L_X$ and SFR, supernova (SN) mechanical energy input rate and $T_X$, which implies a stellar feedback origin of the X-ray emission \citep{Li2013a, Wang2016}. A weaker correlation between $L_X$ and $M_*$ is found in disk galaxies, especially star-burst galaxies, compared to ETGs. 

To characterize such relations at higher redshift, one needs to resort to stacking techniques to overcome the intrinsic dim X-ray emission of the CGM. 
By stacking 250,000 galaxies selected from SDSS with the ROSAT ALL-Sky survey data, \citet{Anderson2015} find the $L_X-M_*$ relation to be steeper than expected from self-similar model predicted above $\log(M_*)>11.0$, implying the existence of non-gravitational heating. 
For the same galaxy sample, a general correlation between the galaxy stellar mass and the Comptonization Y-parameter is measured by \citep{PlanckCollaborationAdeAghanim_2013A&A...557A..52P}. 
With an improved spatial resolution, the $Y$-Mass relation is measured at the high mass end \citep{SchaanFerraroAmodeo_2021PhRvD.103f3513S,PandeyGattiBaxter_2022PhRvD.105l3526P}. 
More recently, \citet{DasChiangMathur_2023ApJ...951..125D} extend the relation down to $\log(M_*)>10.3$ by cross-correlating the WISE and SuperCOSmos photometric redshift galaxy sample \citep{BilickiPeacockJarrett_2016ApJS..225....5B} with ACT and Planck. They find a steeper $Y$-Mass relation than the self-similar model predicted. 

The extended ROentgen Survey with an Imaging Telescope Array (eROSITA) on board the Spektrum Roentgen Gamma (SRG) orbital observatory provides the new X-ray data for stacking experiments. Recently, with eROSITA and its PV/eFEDS observations covering 140 square degrees, \citet{Comparat2022} stack $\sim$16,000 central star-forming and quiescent galaxies selected from the GAMA spectroscopic galaxy survey; later, \citet{Chada2022} stack $\sim$1,600 massive star-forming and quiescent galaxies selected from SDSS in the same eFEDS field. The difference in the sample selection leads to different interpretations of the results (we will discuss in detail in Zhang et al. in prep).

This work uses the state-of-the-art stacking results obtained in \citet{Zhang2024profile} (hereafter Paper I). 
There, we use the first four of the planned eight all-sky surveys (eROSITA all-sky survey, eRASS:4) \citep{Merloni2012,Predehl2021,SunyaevArefievBabyshkin_2021A&A...656A.132S,Merloni2024} and galaxy samples with high completeness \citep{Dey2019, Zou2019,Tinker2021, Tinker2022} to measure the average surface brightness profiles of the CGM, after modeling and subtracting the contaminating X-ray emission from AGN, X-ray Binaries (XRB) and satellite galaxies. In this work, we integrated the X-ray surface brightness profiles in Paper I to calculate the X-ray luminosity. We studied the relations between the X-ray luminosities and galaxies' stellar or halo masses. 

In the literature, the gas within the virial radius is generally called CGM for isolated massive galaxies and is called intragroup medium (IGrM) or intracluster medium (ICM) for more massive systems. However, the separation between them is hard to define physically (and observationally) and is not the goal of this work. 
For simplicity, we define CGM as the gas within a certain radius (i.e., $R_{\rm 500c}$\footnote{$R_{\rm 500c}$ is the radius where the density is 500 times the critical density of the local universe. The halo mass within $R_{\rm 500c}$ is denoted as $M_{\rm 500c}$.}) of the central galaxy, regardless of the halo mass.

The eROSITA X-ray data reduction, stacking method, and galaxy samples are detailed in Paper I. 
This paper is organized as follows. 
The relevant methods and samples are explained in Sect.~\ref{Sec_method}. 
The relations between X-ray luminosity and masses of galaxies are presented in Sect.~\ref{Sec_lxm}.
The implication of the observed scaling relations is discussed in Sect.~\ref{Sec_relations}. 
We used \citet{Planck2020} cosmological parameters: $H_0 = 67.74\ \rm km\,s^{-1}\,Mpc^{-1}$ and $\Omega_{\rm m} = 0.3089$. The $\log$ in this work designates $\log_{10}$. The stellar mass and halo mass in this work are in units of solar mass ($M_\odot$).

In two additional companion papers also based on the measurements from Paper I, we investigate \textit{(i)} the trends as a function of specific star formation rate (Zhang et al. in prep) and \textit{(ii)} the possible dependence on azimuth angle (Zhang et al. in prep).

\section{Methods}
\label{Sec_method}

\subsection{X-ray luminosity}
The total X-ray luminosity $L_{\rm X}$ around a galaxy is calculated by integrating the (background subtracted) surface brightness profiles $S_{\rm X,gal}-S_{\rm X,bg}$ up to a given radius $r$,  
\begin{equation}
    L_{\rm X}= \Sigma_{0}^{r} (S_{\rm X,gal}-S_{\rm X,bg})\times A_{\rm shell}\ [\rm erg\,s^{-1}],
\end{equation}
where $S_{\rm X,gal}$ is the observed rest-frame X-ray surface brightness of stacked galaxies, $S_{\rm X,bg}$ is the background X-ray surface brightness, $A_{\rm shell}$ is the area size of the integrated annulus. 

The uncertainty on $L_{\rm X}$ is estimated from the quadratic sum of the Poisson error and uncertainty estimated with Jackknife re-sampling \citep[][Paper I]{Andraejackknife2010,McIntoshjackknife2016}. 
The Poisson error is negligible, benefiting from the large galaxy sample. The dominant Jackknife uncertainty reflects the scatter of the mean X-ray luminosities of the stacked galaxy population.  

The luminosity is calculated in the energy bin $0.5-2$ keV (in the rest frame). We integrated the X-ray emission within $R_{\rm 500c}$, and the obtained luminosity is denoted as $L_{\rm X,total}$.
Without masking detected X-ray sources, $L_{\rm X,total}$ is the total X-ray luminosity of all sources within $R_{\rm 500c}$, including the point sources (AGN and XRB) and extended CGM. 

We follow the method in Paper I to calculate the CGM luminosity, which we summarize here:

\begin{itemize}
    \item{We mask detected X-ray point sources, integrate the X-ray emission within $R_{\rm 500c}$, correct for the misclassified centrals, and obtain $L_{\rm X,mask}$.}
    \item{We estimate the X-ray emission from unresolved AGN and XRB that reside in the stacked galaxies and the satellite galaxies ($L_{\rm X,AGN+XRB+SAT}$). The XRB luminosity is derived from the stellar mass and SFR of stacked central galaxies according to the empirical model compiled by \citet{Aird2017}\footnote{We verified that using the \citet{Lehmer2019} XRB model (instead of \citet{Aird2017}) does not affect the $L_{\rm X, CGM}$ obtained.}.
    We estimate the unresolved AGN luminosity in central galaxies by stacking optical-AGN-hosting galaxies. The obtained AGN luminosity agrees well with the empirical AGN model compiled by \citet{Comparat2019}. 
    For satellite galaxies, the XRB emission is derived by combining \citet{Aird2017} and the mock SDSS catalog constructed and discussed in Paper I. 
    We neglect the contribution from AGN in satellite galaxies since they have a small occurrence ($<\,10$\%) \citep{Comparat2023}.}
    \item{We subtract $L_{\rm X,AGN+XRB+SAT}$ from $L_{\rm X,mask}$. The residual emission ($L_{\rm X,CGM}$) comes from the hot CGM within $R_{\rm 500c}$:  $L_{\rm X,CGM} = L_{\rm X,mask} - L_{\rm X,AGN+XRB+SAT}$.}   
\end{itemize}

We parametrize the scaling relations between $L_{\rm X,CGM}$ (or $L_{\rm X,total}$) and masses of galaxies by a single power law:
\begin{equation}
\label{eq:singlePL}
    \log(L_{\rm X,CGM})= L_0+\alpha\log(M).
\end{equation}
The single power law has two parameters $L_0$ and the slope $\alpha$. 
We use the maximum likelihood method and the Markov chain Monte Carlo (MCMC) chains to estimate the best-fit parameters and their $1\sigma$ uncertainties \citep[Paper I,][]{emcee2010}.

\subsection{Galaxy samples} \label{Sec_galaxy}

We select galaxies from the SDSS DR7 spectroscopic galaxy catalog ($r_{\rm AB}<17.77$) \citep{Strauss2002}. The spectroscopic redshift ($z_{\rm spec}$) of the galaxy is estimated with an accuracy of $\Delta z_{\rm spec}<10^{-4}$ \citep{Blanton2005}. We limit the maximum redshift of the galaxies, so the sample is approximately volume-limited and complete. The stellar mass of the galaxies is estimated by \citet{Chen2012} and has a typical uncertainty of 0.1 dex. We stack central galaxies only. The central galaxies in SDSS are identified by the halo-based group finder of \citet{Tinker2021}. About 1\% of the central galaxies are misclassified, and we correct it when measuring $L_{\rm X, mask}$. For each group, a halo mass ($M_{\rm 200m}$) is inferred and assigned to the central galaxy\footnote{The halo mass provided in the catalog \citet{Tinker2021} is $M_{\rm 200m}$, the mass within the radius where the mean interior density is 200 times the background universe matter density. To ease the comparison to literature, we convert the $M_{\rm 200m}$ to $M_{\rm 500c}$ with the mass-concentration relation model from \citet{Ishiyama2021}.} \citep{AlpaslanTinker_2020MNRAS.496.5463A,Tinker2021,Tinker2022} .
The \citet{Tinker2022} model reproduces very closely summary statistics of galaxy clustering and galaxy-galaxy lensing \citep{ZehaviZhengWeinberg_2011ApJ...736...59Z,MandelbaumWangZu_2016MNRAS.457.3200M}, meaning that the average halo mass of (large) samples of central galaxies is accurate. 
The SFR of galaxies is estimated by \citet{Brinchmann2004}, which we take to model the XRB emission. 
In this analysis, we do not separate star-forming and quiescent galaxies. 
The galaxy sample selection method is detailed in Paper I. The only difference compared to Paper I's samples is that we split in two the top three stellar and halo mass bins to gain more insight into the scaling relation trend. We build two samples from SDSS DR7:
\begin{itemize}
    \item{\it CEN sample}. This sample includes 85,222 central galaxies in the stellar mass range $10.0<\log(M_*)<11.5$ and spectroscopic redshift $0.01<z_{\rm spec}<0.19$ (see Table~\ref{table:LX:Ms}). 
    \item{\it CEN$_{\rm halo}$ sample}. This sample includes 125,512 central galaxies selected in the halo mass range $\log M_{\rm 200m}=11.5-14.0$ and spectroscopic redshift $0.01<z_{\rm spec}<0.20$ (see Table~\ref{table:LX:Mh}).  
\end{itemize}
By having these two samples, the inversion problem of the stellar-to-halo-mass relation (SHMR) is considered from two complementary angles. 
With the stacks around the CEN and CEN$_{\rm halo}$ samples, we derive the $L_{\rm X,CGM}$.

The deep ($r_{\rm AB}<$ 23.4) DESI Legacy Survey (LS) DR9 allows us to build another larger galaxy sample extended to lower stellar mass and higher redshift \citep{Dey2019}. The photometric redshift of the galaxy is estimated with an accuracy of $\Delta z_{\rm phot} \approx 0.01$, and the stellar mass is derived with typical uncertainties of 0.2 dex \citep{Zou2019}. Albeit the methods to infer the stellar masses are different for the CEN and isolated samples \citep{Chen2012, Zou2019}, we find that the stellar masses of the galaxies in common are consistent within their uncertainties. 
Given the rather large photometric redshift uncertainty, isolated galaxies are selected using angular coordinates only. 
An isolated galaxy's halo does not overlap with any other comparable stellar mass galaxies's halo 
(see details in Paper I):
\begin{itemize}
    \item{\it Isolated sample}. This sample includes 213,514 galaxies in the stellar mass range $9.5<\log(M_*)<11.5$ and photometric redshift $0.01<z_{\rm phot}<0.4$ (see Table~\ref{table:LX:Ms}). The galaxies are isolated galaxies living in under-dense environments.
\end{itemize}
We stack the X-ray around the isolated sample to measure the mean $L_{\rm X,total}$ of galaxies at lower mass and higher redshift than the CEN sample.

\subsection{Simulation datasets and computation of the intrinsic X-ray emission}

To facilitate the interpretation of our observational results, we utilize publicly available simulated datasets from the three distinguished cosmological hydrodynamical simulations: TNG100 from IllustrisTNG \citep{Marinacci2018,Naiman2018,Springel2018,Pillepich2018,Nelson2019a,Nelson2019b}, EAGLE \citep{Crain2015,Schaller2015,Schaye2015,McAlpine2016}, and SIMBA \citep{Dave2019}\footnote{We note that here we use versions of EAGLE and SIMBA that are reprocessed to share the same data format with TNG simulation \citep{Nelson2019a}. The reprocessing ensures that simulated structures, for example, galaxies and their host halos, as well as their properties, are identified in the same manner among the three simulations. This also allows the usage of identical analysis methods.}. These simulations aim to model galaxy formation and evolution within a $\Lambda$CDM universe with consistent cosmological parameters and simulated comoving box ($\sim$100 Mpc per side). The three models, while implementing various astrophysical processes relevant to galaxy formation, differ from each other in terms of modeling feedback from stellar and supermassive black hole activities. These variations on feedback modeling could result in significantly different predictions on the CGM properties \citep{OppenheimerBabulBahe_2021Univ....7..209O,Truong2023,Wright2024}.

We compare the stacking results to simulated galaxies with stellar mass $\log(M_*)>9.5$. The stellar masses of simulated galaxies are measured within twice the half-stellar mass radius. We do not attempt to replicate the selection of the observed galaxy samples, apart from choosing simulated galaxies that are centrals of their host halo.
For each simulated galaxy, the X-ray luminosity of the CGM is measured in the $0.5-2$ keV range and within a radial range of $(0.15-1)\,R_{\rm 500c}$, following the methods detailed in \citet{Truong2021b} and \citet{Comparat2022}. In short, the X-ray luminosity is derived from the thermodynamic properties of gas particles (non-star-forming) by employing a single-temperature APEC model \citep{Smith2001}. The $R_{500c}$ and $M_{500c}$ are the values obtained directly from the simulations output.

We emphasize that the comparisons between observations and simulations in this work are made at face value, as we do not replicate any of the steps involved in the observational inference of the quantities of the studied scaling relations. The computed X-ray emissions of the CGM are intrinsic and do not account for the observational effects, for example, the selection effects introduced by the measurement of the stellar or halo mass. A complete forward model based on hydro-dynamical simulations of the observational process is left for future studies.

\section{Results}\label{Sec_lxm}
This section presents the obtained scaling relations between the X-ray luminosity and $M_*$ or $M_{\rm 500c}$. 
The relationship between $L_{\rm X,total}$ (without masking X-ray sources) and $M_*$ of galaxies is presented in Sect.~\ref{Sec_lxmgal}. 
The inferred $L_{\rm X,CGM}$ to $M_*$ relation compared to the simulations is presented in Sect.~\ref{Sec_lxmcgm}.
The relationships between $L_{\rm X,total}$ or $L_{\rm X,CGM}$ and $M_{\rm 500c}$ of galaxies with comparison to simulations are presented in Sect.~\ref{Sec_lxmh}. 

\subsection{$L_{\rm X,total}-M_*$ relation}\label{Sec_lxmgal}

\begin{table*}[h]
\centering
\caption{Observed rest-frame X-ray luminosity in the 0.5-2 keV band within $R_{\rm 500c}$ in erg s$^{-1}$ of different components for CEN and isolated samples.}

\begin{tabular}{ccccccccccc}
\hline \hline
\multicolumn{2}{c}{$\log_{10}(M_*/M_\odot)$}&\multicolumn{2}{|c|}{redshift}&\multicolumn{2}{|c}{Isolated sample}\\
&&&&(no mask)&(with mask)\\
\hline
9.5&10.0&0.01&0.09&$4.3\pm2.4\times10^{39}$  &$4.2\pm1.2\times10^{39}$\\
10.0&10.5&0.01&0.12&$1.1\pm0.4\times10^{40}$  &$6.0\pm1.5\times10^{39}$ \\
10.5&11.0&0.02&0.17&$2.4\pm0.7\times10^{40}$  &$2.0\pm0.2\times10^{40}$\\
11.0&11.25&0.02&0.33&$1.1\pm0.2\times10^{41}$  &$7.8\pm0.9\times10^{40}$\\
11.25&11.5&0.03&0.40&$4.1\pm0.8\times10^{41}$&$2.8\pm0.3\times10^{41}$\\
\hline
\end{tabular}

\begin{tabular}{ccccccccccc}
\hline \hline
\multicolumn{2}{c}{$\log_{10}(M_*/M_\odot)$}&\multicolumn{2}{|c|}{redshift}&\multicolumn{4}{c}{CEN sample}\\
min&max&min&max&$L_{\rm X,total}$&$L_{\rm X,mask}$&$L_{\rm X,CGM}$& $L_{\rm X,AGN+XRB+SAT}$ \\
&&&&(no mask)&(with mask)&&(with mask)\\
\hline
10.0&10.5&0.01&0.06&$(8.6\pm5.6)\times10^{39}$&$(3.8\pm2.0)\times10^{39}$&$(1.6\pm2.1)\times10^{39}$&$(2.2\pm0.6)\times10^{39}$ \\
10.5&10.75&0.02&0.10&$(2.2\pm0.8)\times10^{40}$&$(1.2\pm0.3)\times10^{40}$&$(6.9\pm3.0)\times10^{39}$&$(5.2\pm1.5)\times10^{39}$ \\
10.75&11.0&0.02&0.10&$(4.2\pm1.0)\times10^{40}$&$(2.1\pm0.4)\times10^{40}$&$(1.3\pm0.4)\times10^{40}$&$(8.2\pm2.4)\times10^{39}$ \\
11.0&11.125&0.02&0.15&$(7.7\pm2.1)\times10^{40}$&$(4.3\pm0.6)\times10^{40}$&$(2.8\pm0.8)\times10^{40}$&$(1.5\pm0.5)\times10^{40}$ \\
11.125&11.25&0.02&0.15&$(1.5\pm0.3)\times10^{41}$&$(8.7\pm0.9)\times10^{40}$&$(6.7\pm1.0)\times10^{40}$&$(2.0\pm0.6)\times10^{40}$ \\
11.25&11.375&0.03&0.19&$(2.9\pm0.5)\times10^{41}$&$(1.8\pm0.2)\times10^{41}$&$(1.6\pm0.2)\times10^{41}$&$(2.5\pm0.7)\times10^{40}$\\
11.375&11.5&0.03&0.19&$(4.8\pm0.4)\times10^{41}$&$(3.7\pm0.3)\times10^{41}$&$(3.4\pm0.3)\times10^{41}$&$(3.4\pm0.9)\times10^{40}$\\
\hline
\end{tabular}
\tablefoot{At different stellar mass bins (first two columns) within each redshift bin (the third and fourth columns), the X-ray luminosity of all X-ray emission ($L_{\rm X, total}$, without masking detected X-ray sources), the X-ray luminosity after excluding detected X-ray point sources ($L_{\rm X, mask}$, with masking X-ray point sources), the X-ray luminosity of hot CGM ($L_{\rm X,CGM}$), and modeled X-ray luminosity from unresolved AGN, XRB sources ($L_{\rm X,AGN+XRB+SAT}$). }
\label{table:LX:Ms}
\end{table*}

\begin{figure}
    \centering
    \includegraphics[width=1.0\columnwidth]{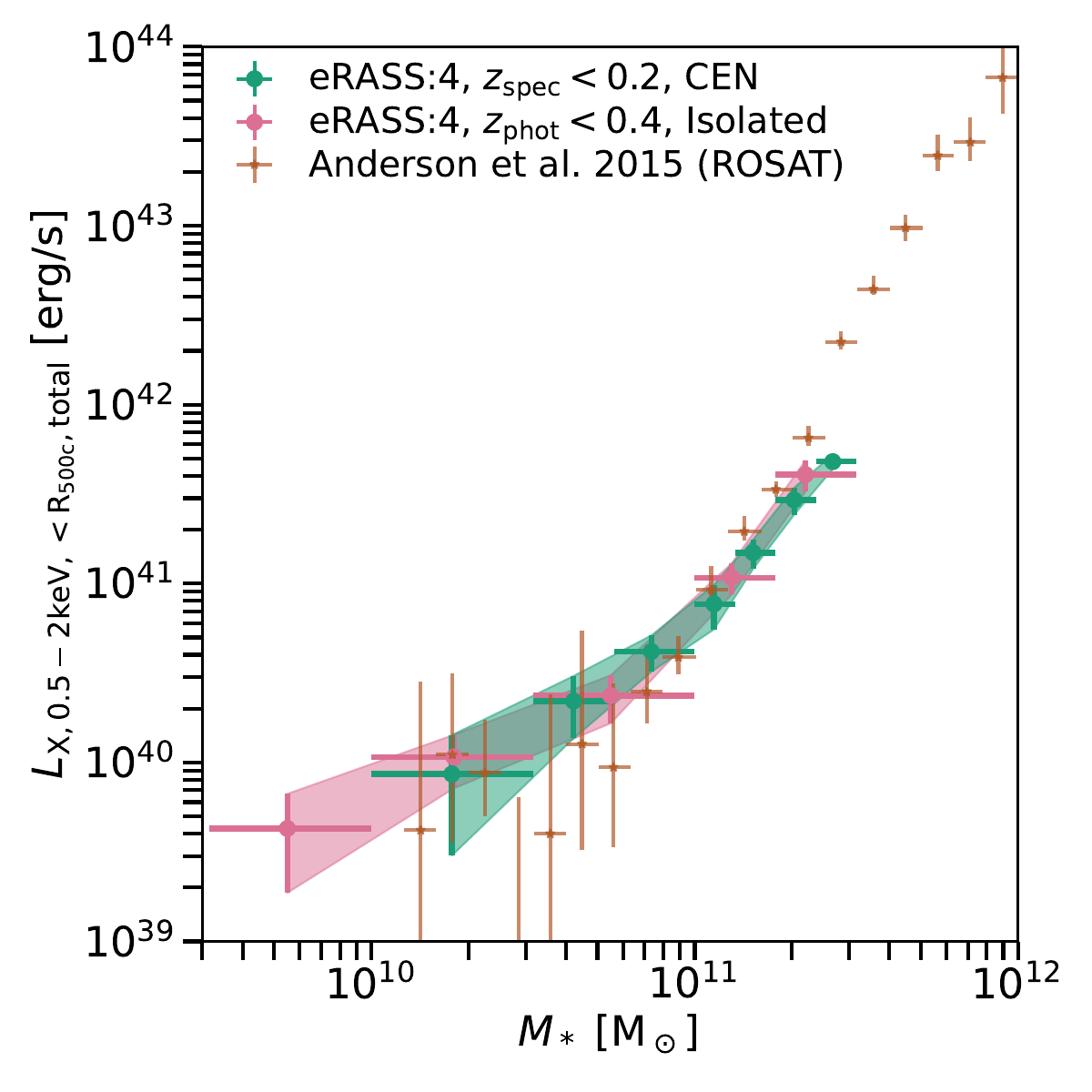}
    \caption{$0.5-2$ keV X-ray luminosity ($L_{\rm X,total}$) within $R_{500c}$ of galaxies without masking detected X-ray sources, as a function of stellar mass $M_*$, for central galaxies (green) within $z_{\rm spec}=0.2$ and isolated galaxies (pink) within $z_{\rm phot}=0.4$. The measurement from \citet{Anderson2015} is plotted in brown. 
    }
        \label{Fig_lxm_gal}
\end{figure}

Based on the central and isolated galaxies stacking, the $L_{\rm X,total}-M_*$ relation is plotted in Fig.~\ref{Fig_lxm_gal}. Note that $L_{\rm X,total}$ includes the emission from AGN, XRB, and hot gas. 
$L_{\rm X,total}$ increases from $(4.3 \pm 2.4)\times10^{39} \rm\, erg\,s^{-1}$ to $(4.8\pm 0.4)\times10^{41} \rm\, erg\,s^{-1}$, when $\log10(M_*)$ increases from $9.5$ to $11.5$ (see Table~\ref{table:LX:Ms}). 
The increase of $\log(L_{\rm X,total})$ with $\log(M_*)$ appears non-linear: below $\log(M_*)\approx 11.0$ the increase is slower, above which it gets faster with $\log(M_*)$.
The $L_{\rm X,total}-M_*$ relations by stacking the isolated and CEN sample give consistent results, with a difference of less than 1$\sigma$. 
Notice that the isolated and CEN samples have a maximum redshift difference of 0.2, which is about a 2 Gyr age difference. We do not detect redshift evolution of the $L_{\rm X,total}-M_*$ relation within the limited redshift range. 

In Fig.~\ref{Fig_lxm_gal}, we also compare our $L_{\rm X,total}-M_*$ relation to \citet[][Fig. 5, top panel]{Anderson2015}. 
They measured the X-ray luminosity of the `locally brightest galaxies' (similar to the CEN sample) using ROSAT. 
The $L_{\rm X,total}-M_*$ relations in the two works agree with a difference less than 1$\sigma$, except for the highest three stellar mass bins. The main reason is the different $M_*$ estimation, that the stellar mass in \citet{Anderson2015} is about $0.1$ dex lower than ours (see the discussions in \citet{Taylor2011} and \citet{Chen2012}). The X-ray data and its reduction, and the galaxy selection are also different in the two works. Full forward models of both measurement processes are needed to reconcile the results at high mass.
Thanks to eRASS:4, significant progress is made: we narrow down the uncertainties on the relation by a factor of about ten at $\log(M_*)=10.7$. With the isolated sample, we measure for the first time the X-ray luminosity of galaxies with a stellar mass of $(3\sim10)\times10^9 M_\odot$ to be $(4.3 \pm 2.4)\times10^{39} \rm erg\,s^{-1}$.

\subsection{$L_{\rm X, CGM}$-$M_*$ relation} \label{Sec_lxmcgm}

\begin{figure}
    \centering
    \includegraphics[width=1.0\columnwidth]{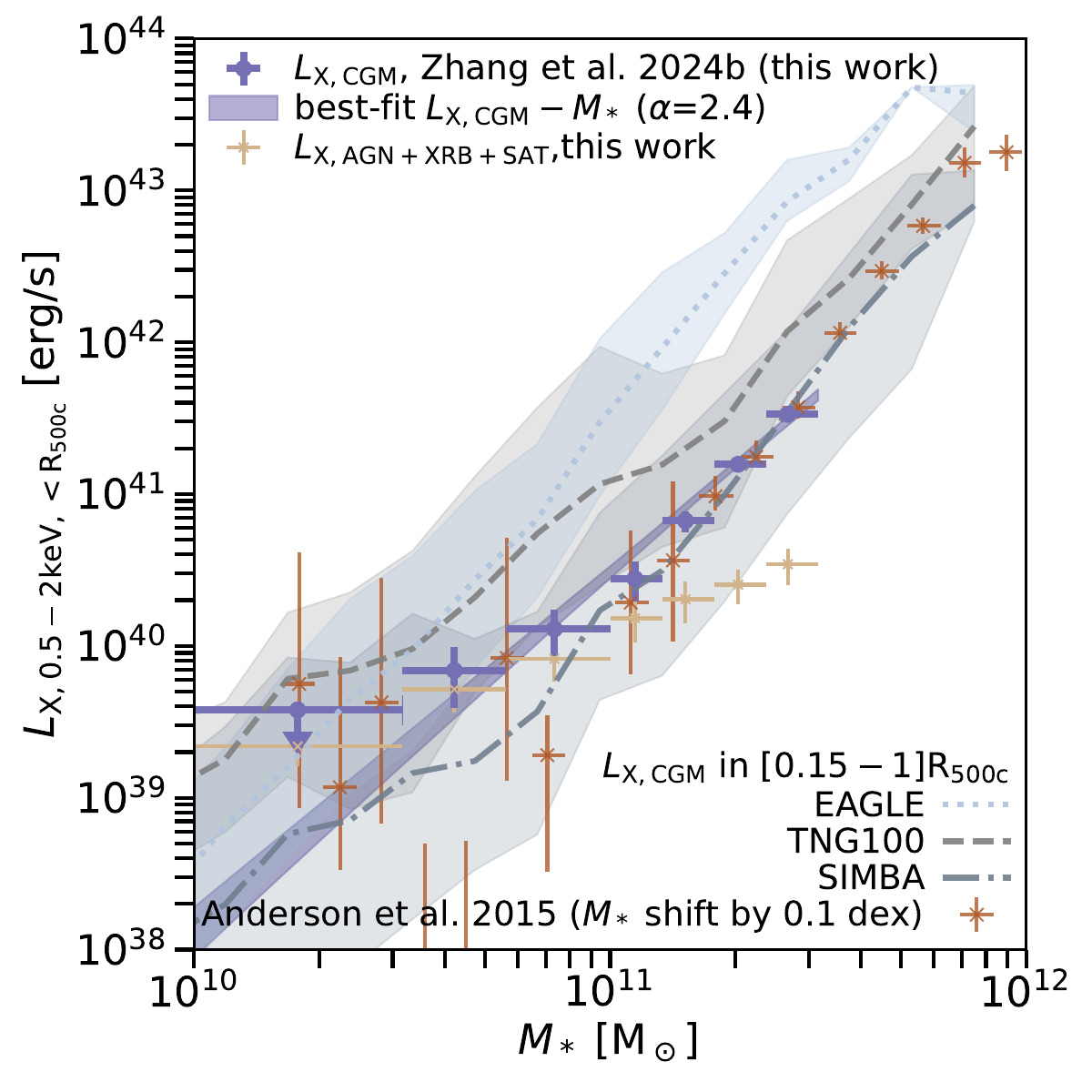}
    \caption{X-ray luminosity vs. stellar mass. 
    The hot CGM X-ray luminosity ($L_{\rm X,CGM}$) measured around the central galaxies (CEN sample) within $R_{500c}$ in $0.5-2$ keV as a function of the $M_*$ is shown in purple. The purple band displays the best-fit single power law model and its $1\sigma$ uncertainty.
    The modeled X-ray luminosity from unresolved AGN, XRB ($L_{\rm X,AGN+XRB+SAT}$) is depicted in tan. 
    The $L_{\rm X,CGM}-M_*$ measurements from \citet{Anderson2015} are plotted with thin brown crosses, notice the $M_*$ is shifted by $0.1$ dex to higher mass to account for the difference in $M_*$ estimation between \citet{Anderson2015} and this work.
    The prediction from the EAGLE, TNG100 and SIMBA simulations compiled following the methodology of \citet{Truong2023} are overplotted with $1\sigma$ uncertainties.}
        \label{Fig_lxm_cgm}
\end{figure}

We present the CGM luminosity $L_{\rm X,CGM}$ and contamination from unresolved sources $L_{\rm X,AGN+XRB+SAT}$ of the CEN sample as a function of $M_*$ in Fig.~\ref{Fig_lxm_cgm}.
We find that the unresolved AGN and XRB contribute to about 60\% of the $L_{\rm X}$ for galaxies with $\log(M_*)=10.0-10.5$ and less than 30\% for galaxies with $\log(M_*)>11.0$. 
The hot CGM thus contributes to about $40$\% of the X-ray luminosity (after masking detected X-ray sources) of the central galaxies at low-mass end and increases to about $>70$\% at high-mass end.
The $L_{\rm X,CGM}$ (purple in Fig.~\ref{Fig_lxm_cgm}, Table ~\ref{table:LX:Ms}) increases from $(1.6 \pm 2.1)\times10^{39} \rm erg\,s^{-1}$ to $(3.4 \pm 0.3)\times10^{41} \rm erg\,s^{-1}$, when $\log(M_*)$ increases from 10.0 to 11.5. 

We compare our CGM luminosity to stellar mass relation ($L_{\rm X,CGM}-M_*$) to the corresponding ones measured by \citet[][Fig. 5, bottom panel]{Anderson2015}. 
To remove the AGN emission, \citet{Anderson2015} define $L_{\rm X,CGM}$ as the X-ray emission in $(0.15-1) \times R_{500}$. This is different from our direct subtraction of emission coming from unresolved AGN obtained by modeling.
Our $L_{\rm X,CGM}-M_*$ relation agrees well with \citet{Anderson2015} after shifting their $M_*$ by $0.1$ dex to higher mass to account for the different $M_*$ estimations. Our measurements significantly improved signal-to-noise.

We fit our measurements with a single power-law (Eq.~ \ref{eq:singlePL}). The $L_{\rm X,CGM}-M_*$ relation below $\log(M_*)=11.5$ can be described with
\begin{equation}\label{Eq_LcgmMs}
    \log L_{\rm X,CGM} = (2.4\pm0.1) \log M_* + (14.6 \pm 1.5).
\end{equation}

We compare the observed $L_{\rm X,CGM}-M_*$ relation to the ones predicted by simulations: EAGLE, TNG1000, and SIMBA. 
The simulations may deviate from the observations at some stellar mass range. Still, given the large scatter, the simulations generally agree with the observations, except for EAGLE (see more discussion in Sect.~\ref{Sec_sim}). 

The baryons follow the gravitational potential well, and the hot CGM luminosity is expected to trace the halo mass \citep{Tumlinson2011}. Thus, by measuring the relation between CGM luminosity and stellar mass, we probe the SHMR in a novel fashion, complementary to probes of SHMR using galaxy clustering and galaxy-galaxy lensing \citep{Leauthaud2012a, VelandervanUitertHoekstra_2014MNRAS.437.2111V, Coupon2015}. 
We discuss the implications of our results further in Sect.~\ref{Sec_lxslxh}. 

\subsection{$L_{\rm X,total}$-$M_{\rm 500c}$ and $L_{\rm X,CGM}$-$M_{\rm 500c}$ relations}
\label{Sec_lxmh}

\begin{table*}[h]
\centering
\caption{Observed rest-frame X-ray luminosity in the 0.5-2 keV band within $R_{\rm 500c}$ in erg s$^{-1}$ of different components for CEN$_{\rm halo}$ sample. }
\begin{tabular}{ccccccccccc}
\hline \hline
\multicolumn{2}{c}{$\log_{10}(M_{\rm 200m})$}&\multicolumn{2}{|c}{$\log_{10}(M_{\rm 500c})$}&\multicolumn{2}{|c|}{redshift}&$L_{\rm X,total}$&$L_{\rm X,mask}$&$L_{\rm X,CGM}$& $L_{\rm X,AGN+XRB+SAT}$\\
min&max&min&max&min&max&(no mask)&(with mask)&&(with mask) \\
&&&&\\
\hline
11.5&12.0&11.3&11.8&0.01&0.08&$(1.2\pm0.5)\times10^{40}$&$(5.2\pm1.5)\times10^{39}$&$(3.0\pm1.6)\times10^{39}$&$(2.2\pm0.5)\times10^{39}$ \\
12.0&12.5&11.8&12.3&0.02&0.13&$(3.9\pm0.7)\times10^{40}$&$(1.9\pm0.3)\times10^{40}$&$(1.1\pm0.4)\times10^{40}$&$(8.3\pm2.2)\times10^{39}$ \\
12.5&12.75&12.3&12.5&0.02&0.16&$(9.7\pm1.6)\times10^{40}$&$(5.0\pm0.9)\times10^{40}$&$(3.4\pm1.0)\times10^{40}$&$(1.6\pm0.6)\times10^{40}$ \\
12.75&13.0&12.5&12.8&0.02&0.16&$(1.5\pm0.3)\times10^{41}$&$(7.0\pm1.7)\times10^{40}$&$(5.7\pm1.7)\times10^{40}$&$(1.3\pm0.4)\times10^{40}$ \\
13.0&13.25&12.8&13.0&0.03&0.20&$(2.6\pm0.8)\times10^{41}$&$(1.5\pm0.2)\times10^{41}$&$(1.3\pm0.2)\times10^{41}$&$(1.6\pm0.4)\times10^{40}$\\
13.25&13.5&13.0&13.3&0.03&0.20&$(4.5\pm0.6)\times10^{41}$&$(3.7\pm0.4)\times10^{41}$&$(3.5\pm0.4)\times10^{41}$&$(2.3\pm0.6)\times10^{40}$\\
13.5&13.75&13.3&13.5&0.03&0.20&$(8.6\pm1.0)\times10^{41}$&$(6.4\pm0.7)\times10^{41}$&$(6.1\pm0.7)\times10^{41}$&$(3.2\pm0.7)\times10^{40}$\\
13.75&14.0&13.5&13.7&0.03&0.20&$(1.6\pm0.2)\times10^{42}$&$(1.4\pm0.1)\times10^{42}$&$(1.3\pm0.1)\times10^{42}$&$(4.6\pm1.0)\times10^{40}$\\
\hline
\end{tabular}
\tablefoot{The CEN$_{\rm halo}$ sample binned in $M_{\rm 200m}$ derived based on the group finder algorithm \citep{Tinker2021}, the corresponding $M_{\rm 500c}$ derived considering the concentration model from \citet{Ishiyama2021} and the redshift bins. The luminosities are defined in Table~\ref{table:LX:Ms}.}
\label{table:LX:Mh}
\end{table*}

\begin{figure}[h]
    \centering
    \includegraphics[width=1.0\columnwidth]{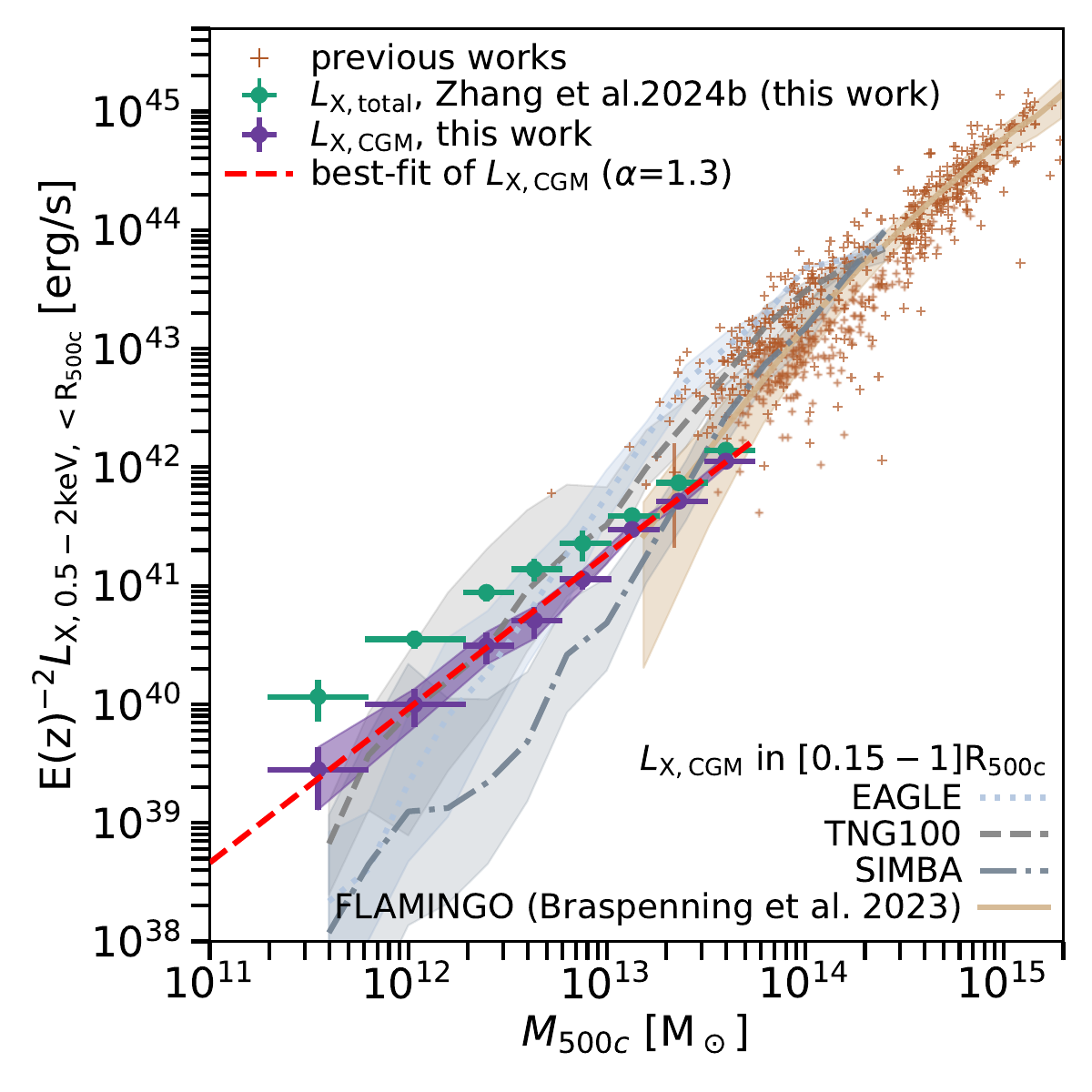}
    \caption{Total X-ray luminosity ($L_{\rm X,total}$, green) and X-ray luminosity of hot gas ($L_{\rm X,CGM}$, purple) within $R_{\rm 500c}$ in $0.5-2$ keV as a function of the $M_{\rm 500c}$. 
    The red dashed line is the best-fit single power law to our measurements of the $L_{\rm X,CGM}-M_{\rm 500c}$ relation. 
    The brown data points are taken from \citet{Eckmiller2011,Lovisari2015,Mantz2016,Schellenberger2017,Andreon2017,Adami2018,Bulbul2019,Lovisari2020,Liu2022,Popesso2024} (Notice the presence of selection effects and Malmquist bias in these observations). 
    The $L_{\rm X,CGM}-M_{\rm 500c}$ relations predicted by simulations EAGLE, TNG100, SIMBA and FLAMINGO with $1\sigma$ uncertainties are overplotted \citep{Braspenning2023}.}
        \label{Fig_lxm_halo}
\end{figure}

This section presents the stack of the CEN$_{\rm halo}$ sample. 
The $L_{\rm X,total}-M_{\rm 500c}$ relation obtained is plotted in Fig.~\ref{Fig_lxm_halo} in green and listed in Table~\ref{table:LX:Mh}. We scale $L_{\rm X,total}$ and $L_{\rm X,CGM}$ by $E(z)^{-2}$ considering a self-similar redshift evolution\footnote{The $E(z)$ is imported to account for the redshift evolution of quantities related to the overdensity (i.e., $R_{\rm 500c}$).The self-similar redshift evolution of $E(z)^{-2}$ works for the massive clusters, where the X-ray emission from the bremsstrahlung process dominates. For lower mass structures where the line emission dominates, $E(z)^{-5/3}$ is usually taken \citep{Lovisari2022}. Considering the CEN$_{\rm halo}$ sample is below $z=0.2$ ($E(z)\approx 1.0-1.1$) and to keep consistency with clusters studies, we adopt a $E(z)^{-2}$ scaling throughout.}. 

The $L_{\rm X,total}-M_{\rm 500c}$ relation is measured down to $\log(M_{\rm 500})=11.3$. The $L_{\rm X,total}$ increases from $(1.2 \pm 0.5)\times10^{40}\ \rm erg\,s^{-1}$ at $\log(M_{\rm 500c})=11.3$ to $(1.6 \pm 0.2)\times10^{42}\ \rm erg\,s^{-1}$ at $\log(M_{\rm 500c})=13.7$.
The $\log L_{\rm X,total} - \log M_{\rm 500c}$ relation is quasi-linear and can be described with 
\begin{equation}\label{Eq_LtMh}
    \log L_{\rm X,total} = (1.02\pm0.04)\log M_{\rm 500c} +(28.3 \pm 0.5).
\end{equation}

The X-ray luminosity of the hot CGM within $R_{500c}$ ($L_{\rm X,CGM}$) as a function of $M_{\rm 500c}$ is also plotted in Fig.~\ref{Fig_lxm_halo} (purple data points) and listed in Table~\ref{table:LX:Mh}. 
We find that $L_{\rm X,CGM}$ increases with $M_{\rm 500c}$ from $(3.0 \pm 1.6)\times10^{39}\ \rm erg\,s^{-1}$ at $\log(M_{\rm 500c})=11.3$ to $(1.3 \pm 0.1)\times10^{42}\ \rm erg\,s^{-1}$ at $\log(M_{\rm 500c})=13.7$. 
We fit the $L_{\rm X,CGM}-M_{500c}$ relation with Eq.~\ref{eq:singlePL}, and get 
\begin{equation}\label{Eq_LcgmMh}
    \log(L_{\rm X,CGM})=(1.32\pm0.05)\log(M_{500c})+(24.1 \pm 0.7).
\end{equation}
The $L_{\rm X,CGM}-M_{\rm 500c}$ relation has been extensively studied at the high-mass end ($\log(M_{\rm 500c})>14.0$) \citep{Eckmiller2011,Lovisari2015,Mantz2016,Schellenberger2017,Andreon2017,Adami2018,Bulbul2019,Lovisari2020,Chiu2022}.
By stacking the galaxy groups from the GAMA survey, the X-ray emission is detected down to $\log(M_{\rm 500c})=13.5$ \citep{Popesso2024}. 
These previous works are compared to our result in Fig.~\ref{Fig_lxm_halo}. Our measurement extends the $L_{\rm X,CGM}-M_{500c}$ relation down to lower mass structure and narrows down the uncertainty by a factor of $>10$ for $\log(M_{\rm 500c})<13.5$ structures. 
Due to the selection effects that only X-ray-detectable galaxy groups and galaxy clusters are measured in the previous works (except \citet{Popesso2024}), the previously measured $L_{\rm X,CGM}$ appears brighter than our stacking analysis at $\log(M_{\rm 500c})=13.0-13.7$. 

With consideration of the selection function of galaxy clusters and uncertainty on the $M_{\rm 500c}$, the slope of $L_{\rm X,CGM}-M_{500c}$ ranges between $1.4-2.0$ \citep{Lovisari2020,Chiu2022}.
Our observed $L_{\rm X,CGM}-M_{500c}$ relation at $\log(M_{\rm 500c})<14.0$ has a shallower slope than the literature measurements at $\log(M_{\rm 500c})>14.0$. 
Though the same hot medium within $R_{\rm 500c}$ is measured, the methodologies differ between ours and the literature. In the literature, the luminosity is derived from spectra fitting and the X-ray profile for each galaxy cluster selected in the X-ray or millimeter wavelength bands \citep{Eckmiller2011,Bulbul2019}. 
We measure an average luminosity for a complete central galaxy sample. 
Further investigation is necessary to figure out the difference caused by the methodology.
The shallower slope could be related to the influence of feedback processes, which might be different at low-mass and high-mass (see more discussion in Sect. ~\ref{Sec_similarity}).
We discuss the comparison with simulations in Sect.~\ref{Sec_sim}.  

\section{Discussion} \label{Sec_relations}

\subsection{Comparing $L_{\rm X,CGM}-M_{500c}$ to self-similar model} \label{Sec_similarity}

Under the assumption that the gas is gravitationally heated only, and the systems (galaxies, galaxy groups, galaxy clusters) reached equilibrium and are isothermal, the properties of the hot gas scale with the dark matter halo mass following the self-similar model \citep{Kaiser1986,Kravtsov2012,Lovisari2021}. The deviation of observed relations from the self-similar model implies that extra non-gravitational mechanisms affect the gas's heating, cooling, or ejection. 

Following \citet{Lovisari2021}, to make the comparison to the scaling relations predicted from the self-similar model easier, in this section, we name the systems with $\log(M_{\rm 500c})>13.4$ as galaxy clusters, $\log(M_{\rm 500c})=12.8-13.4$ as galaxy groups, $\log(M_{\rm 500c})<12.8$ as galaxies. The self-similar relation of $L_{\rm X,CGM}-M_{500c}$ of galaxy clusters has an index of about 0.78 (solar metallicity gas, $Z=Z_\odot$) to 0.94 ($Z=0.3 Z_\odot$) \citep{Lovisari2021}. 
The observed slope of the $L_{\rm X,gas}-M_{500c}$ relation above $\log(M_{\rm 500c})=14.0$ ranges between $1.4-2.0$, steeper than self-similar prediction \citep{Lovisari2020, Chiu2022}. 
This has given rise to a widespread discussion of AGN feedback models and their role in heating/modifying the ICM \citep[e.g.,][]{Fabian2012,LeBrun2014,EckertGaspariGastaldello_2021Univ....7..142E,OppenheimerBabulBahe_2021Univ....7..209O, Pop2022,Schaye2023}.

For lower-mass systems, the self-similar relation of $L_{\rm X,gas}-M_{500c}$ is expected to change due to the complexity of the emissivity of the hot plasma as a function of temperature and metallicity in the 0.5-2 keV band \citep{Lovisari2021}. 
For galaxy groups, the index ranges from 0 ($Z=Z_\odot$) to 0.6 ($Z=0.3 Z_\odot$). However, our observed $L_{\rm X,gas}-M_{\rm 500c}$ has an index of $\approx 1.32$, which is much steeper. It strengthens further the need for non-gravitational processes, probably linked to AGN feedback, to explain the properties of the hot CGM at the galaxy group scale \citep{EckertGaspariGastaldello_2021Univ....7..142E,Bahar2024}. 

For galaxies, the self-similar model index ranges from 1.5 ($Z=Z_\odot$) to 1.6 ($Z=0.3 Z_\odot$) \citep{Lovisari2021}. Here, we obtain an index of $\approx 1.33$, which is about 11--17\% shallower than the self-similar prediction. The two indexes' closeness might imply that these systems' non-gravitational processes (for example, AGN and stellar feedback) are sub-dominant. 
Consistently, the hot CGM around galaxies with $\log(M_{\rm 200m})>11.5$ and redshift $z<0.5$ will reach equilibrium and virialization \citep{Stern2021,Faucher2023}.
In the future, we expect that X-ray spectral fitting will constitute a valuable source of information to gain insights into the physical processes behind this observed luminosity, for example, line cooling, Bremsstrahlung, and metallicity.

\subsection{Comparison to cosmological galaxy simulations} \label{Sec_sim}

The discussion in Sect.~\ref{Sec_similarity} emphasizes that the observed scaling relations contain information about feedback processes. 
In Fig.~\ref{Fig_lxm_cgm} and Fig.~\ref{Fig_lxm_halo}, we compare the observed $L_{\rm X,CGM}-M_{*}$ and $L_{\rm X,CGM}-M_{500c}$ relations to the EAGLE, TNG100, and SIMBA simulations \citep[as compiled by][]{Truong2023}.  In Fig.~\ref{Fig_lxm_halo}, we also compare to the output of the FLAMINGO \citep{Braspenning2023} simulation for halo masses above the group-mass scale.

For the $L_{\rm X,CGM}-M_*$ relation (Fig.~\ref{Fig_lxm_cgm}), the three simulations EAGLE, TNG100, and SIMBA show considerable model-to-model variation across the considered stellar mass range, in addition to approximately a 1-dex galaxy-to-galaxy variation across most of the mass range. Among the three simulations, SIMBA predicts the lowest X-ray emissions due to a smaller amount of hot gas contained in the CGM than the other two simulations (\citealt{Truong2023}). Both SIMBA and TNG mean CGM X-ray luminosities are within 1 dex of the observation, whereas for EAGLE only at galaxy stellar masses $\log(M_*)={10.0-11.0}$. For galaxies with stellar masses above $\log(M_*)=11.0$, EAGLE significantly overpredicts the CGM emissions compared to the observed results. This discrepancy suggests that the AGN feedback model implemented in EAGLE may be less efficient at removing gas from massive haloes \citep{Davies2019}.

The $L_{\rm X,CGM}-M_{500c}$ (Fig.~\ref{Fig_lxm_halo}) relations predicted by the simulations agree with the observations within $1\sigma$ down to $\log(M_{\rm 500c})=12.0$, with a slight deviation from SIMBA at lower mass.
All simulation predictions seem generally steeper than the observation, though within scatter.
According to this comparison, and differently from Fig.~\ref{Fig_lxm_cgm}, EAGLE does not overpredict the $L_{\rm X, CGM}$ over 1-dex, but SIMBA appears to underpredict the CGM luminosity at fixed halo mass. 
It is important to note that the $M_{500c}-M_{*}$ relationships in the simulations are not exactly the same as the one adopted in \citet{Tinker2021} \citep{MandelbaumWangZu_2016MNRAS.457.3200M,Wright2024}. This complicates the interpretation of the possible discrepancies between simulation predictions and observational data, as differences in the $L_{X}-M_{500c}$ relationship could arise from both the underlying feedback physics and the methodologies used to estimate $M_{500c}$ and stellar mass. 


More detailed comparisons are required to draw quantitative and decisive conclusions on the nature of the feedback, the gas enrichment process, and the heating mechanism {in current cosmological simulations of galaxies}. In this work, we do not create a dedicated mock catalog mimicking possible selection effects, which could improve the accuracy of the comparison. 
If a hydro-dynamical simulation reproduces both relations between halo mass, stellar mass and the CGM or the total (adding AGN and XRB) luminosity, it would constitute good evidence for the feedback mechanisms to be accurate. This, however, also requires accurate prediction of the X-ray emission from AGN and XRB \citep[e.g.,][]{Biffi2018MNRAS.481.2213B, Vladutescu-Zopp2023AA...669A..34V,Kyritsis2024}.

\subsection{The connection between $L_{\rm X,CGM}-M_{*}$ and $L_{\rm X,CGM}-M_{\rm 500c}$ relations} \label{Sec_lxslxh}

\begin{figure}
    \centering
    \includegraphics[width=1.0\columnwidth]{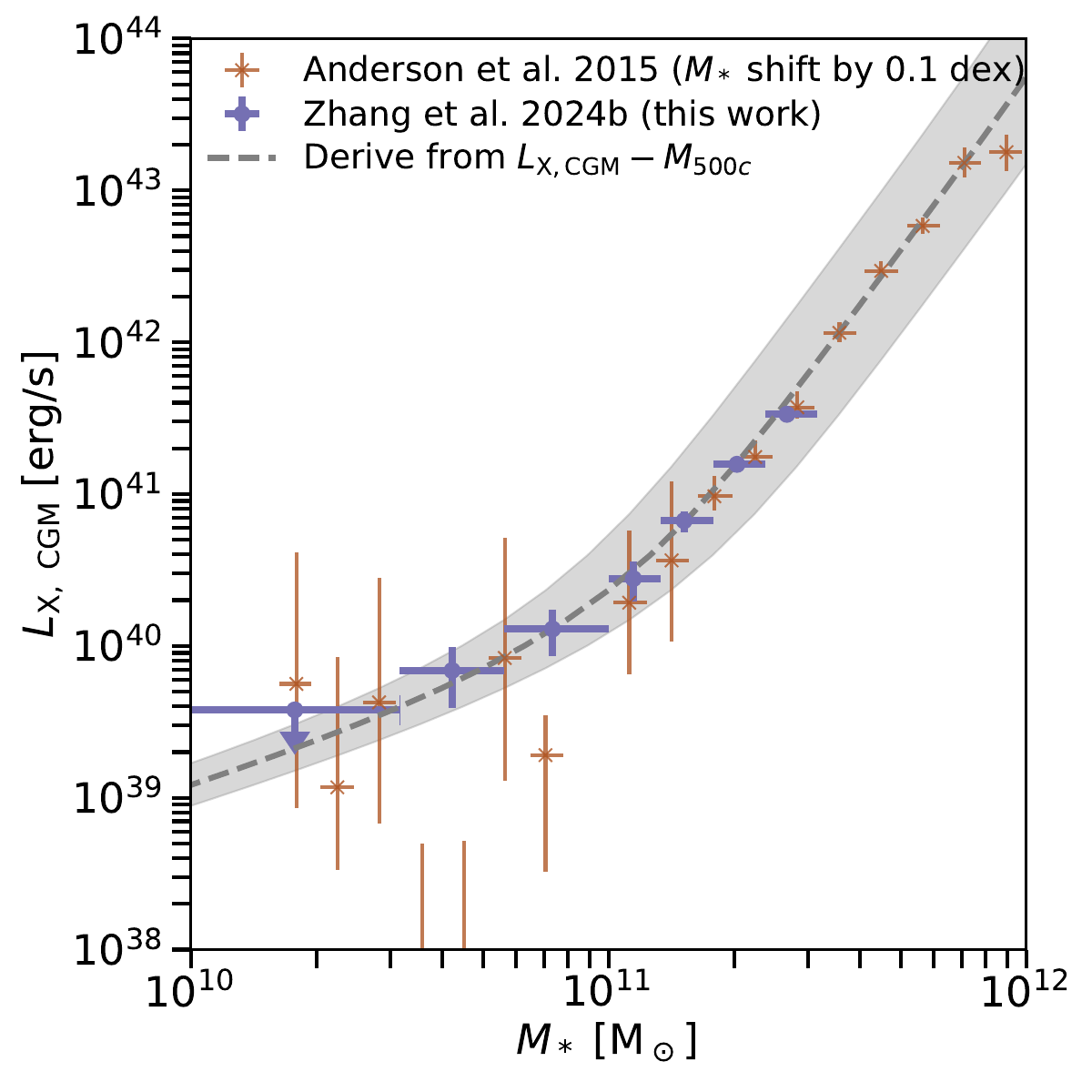}
    \caption{Observed X-ray luminosity of CGM as a function to $M_*$ (purple, brown points), compared to the relation derived by convolving the observed $L_{\rm X,CGM}-M_{\rm 500c}$ relation with the stellar-to-halo-mass relation (gray band). The $L_{\rm X,CGM}-M_*$ measurements from \citet{Anderson2015} is shifted by $0.1$ dex to higher mass to account for the difference in $M_*$ estimation between \citet{Anderson2015} and this work.}
        \label{Fig_lhls}
\end{figure}

The SHMR of the central galaxies selected by \citet{Tinker2021} is well represented by the broken power law 
\begin{equation}\label{Eq_SHMR_tinker}
    M_*= 0.07 M_{\rm 200m} \left[\left( \frac{M_{\rm 200m}}{1.3\times10^{12}}\right)^{-0.6}+\left(\frac{M_{\rm 200m}}{1.3\times10^{12}}\right)^{0.7}\right]^{-1},
\end{equation}
with a scatter of about 0.15 dex.
We then convolve the SHMR to $L_{\rm X,CGM}-M_{\rm 500c}$ to predict the relation between $L_{\rm X,CGM}$ and $M_*$. In Fig.~\ref{Fig_lhls}, we show both the observed and predicted (from SHMR and $L_{\rm X,CGM}-M_{\rm 500c}$) $L_{\rm X,CGM}-M_{*}$ relations. The agreement of the two indeed demonstrates that the measured hot X-ray (CGM) luminosity is a good tracer of the gravitation potential of the halos.
This brings the possibility of a new method to measure the SHMR relation by jointly fitting the relations $L_{\rm X,CGM}-M_{*}$ and $L_{\rm X,CGM}-M_{\rm 500c}$. 

\section{Summary and conclusions}

In this work, we provide an updated and extended quantification of the scaling relations between the X-ray emission from the hot circumgalactic medium (CGM) and two main galaxy properties, stellar and halo mass. These scaling relations can be used to constrain galaxy evolution models. 

By following up on the analysis presented in Paper I \citep{Zhang2024profile}, we extracted the integrated total luminosities within a fixed aperture of $R_{\rm 500c}$ directly measured by stacking eRASS:4 X-ray photons around galaxies in well-defined, highly complete samples built from the SDSS and LS DR9 galaxy catalogs that contain 85222 central galaxies split into stellar mass bins, 125512 central galaxies split in halo mass bins, and 213514 isolated galaxies.
We report the following three findings, which significantly extend our understanding of the hot CGM around galaxies and constitute a new benchmark to constrain galaxy formation and evolution models:
   \begin{itemize}
      \item The scaling relation between the X-ray luminosity $L_{\rm X,total}$ (total X-ray emission within $R_{\rm 500c}$) and the stellar mass of galaxies ($L_{\rm X,total}-M_*$) is measured for galaxies with $\log(M_*)>9.5$ thanks to the large photometric isolated galaxy sample (Fig.~\ref{Fig_lxm_gal}).
      \item The scaling relation between the X-ray luminosity of the hot CGM and the stellar mass ($L_{\rm X,CGM}-M_*$) is measured for central galaxies with $\log(M_*)>10.0$ (Fig.~\ref{Fig_lxm_cgm}). A single power law with indice of $2.4\pm0.1$ describes the $L_{\rm X,CGM}-M_*$ relation (Eq.~\ref{Eq_LcgmMs}).
      \item The scaling relations between total X-ray luminosity and halo mass of galaxies ($L_{\rm X,total}-M_{\rm 500c}$); X-ray luminosity of hot CGM and halo mass ($L_{\rm X,CGM}-M_{500c}$) are measured down to $M_{\rm 500c}=11.3$ 
      (Fig.~\ref{Fig_lxm_halo}). In this mass range, the $L_{\rm X,CGM}-M_{500c}$ relation can be described by a single power law with index $1.32\pm0.05$ (Eq.~\ref{Eq_LcgmMh}). 
   \end{itemize}

By comparing the measured $L_{\rm X,CGM}-M_{500c}$ scaling relation to the self-similar model, we demonstrated the necessity of non-gravitational processes at the galaxy group scale, while these processes appear sub-dominant at the galaxy scale. Additionally, comparisons with the IllustrisTNG, EAGLE, and SIMBA simulations reveal that current cosmological galaxy simulations broadly agree with our observational results, although significant deviations are evident in specific mass ranges.

In the future, with deeper spectroscopic galaxy surveys DESI BGS and 4MOST \citep{Zarrouk2022,HahnWilsonRuiz-Macias_2023AJ....165..253H,deJongAgertzBerbel_2019Msngr.175....3D, FinoguenovMerloniComparat_2019Msngr.175...39F}, we will enlarge the spectroscopic galaxy sample to cover wider areas (3-4 times that used here) and extend further in redshift (up to $z<0.4$).
We expect an apparent increase in signal-to-noise and to obtain a detailed view of the hot CGM and possible relation to intermediate-mass black holes in and around dwarf galaxies ($\log(M_*)<10$) \citep{Arcodia2024}. 

\begin{acknowledgements}
We thank the anonymous referee for thoughtful comments
that improved the manuscript.
This project acknowledges financial support from the European Research Council (ERC) under the European Union's Horizon 2020 research and innovation program HotMilk (grant agreement No. 865637). GP acknowledges support from Bando per il Finanziamento della Ricerca Fondamentale 2022 dell'Istituto Nazionale di Astrofisica (INAF): GO Large program and from the Framework per l'Attrazione e il Rafforzamento delle Eccellenze (FARE) per la ricerca in Italia (R20L5S39T9). NT acknowledges support from NASA under award number 80GSFC21M0002. PP has received funding from the European Research Council (ERC) under the European Union's Horizon Europe research and innovation program ERC CoG (Grant agreement No. 101045437).\\
This work is based on data from eROSITA, the soft X-ray instrument aboard SRG, a joint Russian-German science mission supported by the Russian Space Agency (Roskosmos), in the interests of the Russian Academy of Sciences represented by its Space Research Institute (IKI), and the Deutsches Zentrum für Luft- und Raumfahrt (DLR). The SRG spacecraft was built by Lavochkin Association (NPOL) and its subcontractors, and is operated by NPOL with support from the Max Planck Institute for Extraterrestrial Physics (MPE).\\

The development and construction of the eROSITA X-ray instrument was led by MPE, with contributions from the Dr. Karl Remeis Observatory Bamberg \& ECAP (FAU Erlangen-Nuernberg), the University of Hamburg Observatory, the Leibniz Institute for Astrophysics Potsdam (AIP), and the Institute for Astronomy and Astrophysics of the University of Tübingen, with the support of DLR and the Max Planck Society. The Argelander Institute for Astronomy of the University of Bonn and the Ludwig Maximilians Universität Munich also participated in the science preparation for eROSITA.\\
The eROSITA data shown here were processed using the eSASS/NRTA software system developed by the German eROSITA consortium. \\
Funding for the SDSS and SDSS-II has been provided by the Alfred P. Sloan Foundation, the Participating Institutions, the National Science Foundation, the U.S. Department of Energy, the National Aeronautics and Space Administration, the Japanese Monbukagakusho, the Max Planck Society, and the Higher Education Funding Council for England. The SDSS Web Site is http://www.sdss.org/.

The SDSS is managed by the Astrophysical Research Consortium for the Participating Institutions. The Participating Institutions are the American Museum of Natural History, Astrophysical Institute Potsdam, University of Basel, University of Cambridge, Case Western Reserve University, University of Chicago, Drexel University, Fermilab, the Institute for Advanced Study, the Japan Participation Group, Johns Hopkins University, the Joint Institute for Nuclear Astrophysics, the Kavli Institute for Particle Astrophysics and Cosmology, the Korean Scientist Group, the Chinese Academy of Sciences (LAMOST), Los Alamos National Laboratory, the Max-Planck-Institute for Astronomy (MPIA), the Max-Planck-Institute for Astrophysics (MPA), New Mexico State University, Ohio State University, University of Pittsburgh, University of Portsmouth, Princeton University, the United States Naval Observatory, and the University of Washington.
\\
The DESI Legacy Imaging Surveys consist of three individual and complementary projects: the Dark Energy Camera Legacy Survey (DECaLS), the Beijing-Arizona Sky Survey (BASS), and the Mayall z-band Legacy Survey (MzLS). DECaLS, BASS and MzLS together include data obtained, respectively, at the Blanco telescope, Cerro Tololo Inter-American Observatory, NSF’s NOIRLab; the Bok telescope, Steward Observatory, University of Arizona; and the Mayall telescope, Kitt Peak National Observatory, NOIRLab. NOIRLab is operated by the Association of Universities for Research in Astronomy (AURA) under a cooperative agreement with the National Science Foundation. Pipeline processing and analyses of the data were supported by NOIRLab and the Lawrence Berkeley National Laboratory (LBNL). Legacy Surveys also uses data products from the Near-Earth Object Wide-field Infrared Survey Explorer (NEOWISE), a project of the Jet Propulsion Laboratory/California Institute of Technology, funded by the National Aeronautics and Space Administration. Legacy Surveys was supported by: the Director, Office of Science, Office of High Energy Physics of the U.S. Department of Energy; the National Energy Research Scientific Computing Center, a DOE Office of Science User Facility; the U.S. National Science Foundation, Division of Astronomical Sciences; the National Astronomical Observatories of China, the Chinese Academy of Sciences and the Chinese National Natural Science Foundation. LBNL is managed by the Regents of the University of California under contract to the U.S. Department of Energy. The complete acknowledgments can be found at https://www.legacysurvey.org/acknowledgment/.\\
The Siena Galaxy Atlas was made possible by funding support from the U.S. Department of Energy, Office of Science, Office of High Energy Physics under Award Number DE-SC0020086 and from the National Science Foundation under grant AST-1616414.\\

\end{acknowledgements}

\bibliographystyle{aa}
\bibliography{ref.bib}

\begin{thebibliography}{101}
\expandafter\ifx\csname natexlab\endcsname\relax\def\natexlab#1{#1}\fi

\bibitem[{{Adami} {et~al.}(2018){Adami}, {Giles}, {Koulouridis}, {Pacaud},
  {Caretta}, {Pierre}, {Eckert}, {Ramos-Ceja}, {Gastaldello}, {Fotopoulou},
  {Guglielmo}, {Lidman}, {Sadibekova}, {Iovino}, {Maughan}, {Chiappetti},
  {Alis}, {Altieri}, {Baldry}, {Bottini}, {Birkinshaw}, {Bremer}, {Brown},
  {Cucciati}, {Driver}, {Elmer}, {Ettori}, {Evrard}, {Faccioli}, {Granett},
  {Grootes}, {Guzzo}, {Hopkins}, {Horellou}, {Lef{\`e}vre}, {Liske}, {Malek},
  {Marulli}, {Maurogordato}, {Owers}, {Paltani}, {Poggianti}, {Polletta},
  {Plionis}, {Pollo}, {Pompei}, {Ponman}, {Rapetti}, {Ricci}, {Robotham},
  {Tuffs}, {Tasca}, {Valtchanov}, {Vergani}, {Wagner}, {Willis}, \& {XXL
  Consortium}}]{Adami2018}
{Adami}, C., {Giles}, P., {Koulouridis}, E., {et~al.} 2018, \aap, 620, A5

\bibitem[{{Aird} {et~al.}(2017){Aird}, {Coil}, \& {Georgakakis}}]{Aird2017}
{Aird}, J., {Coil}, A.~L., \& {Georgakakis}, A. 2017, \mnras, 465, 3390

\bibitem[{{Alpaslan} \& {Tinker}(2020)}]{AlpaslanTinker_2020MNRAS.496.5463A}
{Alpaslan}, M. \& {Tinker}, J.~L. 2020, \mnras, 496, 5463

\bibitem[{{Anderson} {et~al.}(2015){Anderson}, {Gaspari}, {White}, {Wang}, \&
  {Dai}}]{Anderson2015}
{Anderson}, M.~E., {Gaspari}, M., {White}, S. D.~M., {Wang}, W., \& {Dai}, X.
  2015, \mnras, 449, 3806

\bibitem[{{Andrae}(2010)}]{Andraejackknife2010}
{Andrae}, R. 2010, arXiv e-prints, arXiv:1009.2755

\bibitem[{{Andreon} {et~al.}(2017){Andreon}, {Wang}, {Trinchieri}, {Moretti},
  \& {Serra}}]{Andreon2017}
{Andreon}, S., {Wang}, J., {Trinchieri}, G., {Moretti}, A., \& {Serra}, A.~L.
  2017, \aap, 606, A24

\bibitem[{{Arcodia} {et~al.}(2024){Arcodia}, {Merloni}, {Comparat}, {Dwelly},
  {Seppi}, {Zhang}, {Buchner}, {Georgakakis}, {Haberl}, {Igo}, {Kyritsis},
  {Liu}, {Nandra}, {Ni}, {Ponti}, {Salvato}, {Ward}, {Wolf}, \&
  {Zezas}}]{Arcodia2024}
{Arcodia}, R., {Merloni}, A., {Comparat}, J., {et~al.} 2024, \aap, 681, A97

\bibitem[{{Bahar} {et~al.}(2024){Bahar}, {Bulbul}, {Ghirardini}, {Sanders},
  {Zhang}, {Liu}, {Clerc}, {Artis}, {Balzer}, {Biffi}, {Bose}, {Comparat},
  {Dolag}, {Garrel}, {Hadzhiyska}, {Hern{\'a}ndez-Aguayo}, {Hernquist},
  {Kluge}, {Krippendorf}, {Merloni}, {Nandra}, {Pakmor}, {Popesso},
  {Ramos-Ceja}, {Seppi}, {Springel}, {Weller}, \& {Zelmer}}]{Bahar2024}
{Bahar}, Y.~E., {Bulbul}, E., {Ghirardini}, V., {et~al.} 2024, arXiv e-prints,
  arXiv:2401.17276

\bibitem[{{Biffi} {et~al.}(2018){Biffi}, {Dolag}, \&
  {Merloni}}]{Biffi2018MNRAS.481.2213B}
{Biffi}, V., {Dolag}, K., \& {Merloni}, A. 2018, \mnras, 481, 2213

\bibitem[{{Bilicki} {et~al.}(2016){Bilicki}, {Peacock}, {Jarrett}, {Cluver},
  {Maddox}, {Brown}, {Taylor}, {Hambly}, {Solarz}, {Holwerda}, \&
  et~al.}]{BilickiPeacockJarrett_2016ApJS..225....5B}
{Bilicki}, M., {Peacock}, J.~A., {Jarrett}, T.~H., {et~al.} 2016, \apjs, 225, 5

\bibitem[{{Blanton} {et~al.}(2005){Blanton}, {Schlegel}, {Strauss},
  {Brinkmann}, {Finkbeiner}, {Fukugita}, {Gunn}, {Hogg}, {Ivezi{\'c}}, {Knapp},
  {Lupton}, {Munn}, {Schneider}, {Tegmark}, \& {Zehavi}}]{Blanton2005}
{Blanton}, M.~R., {Schlegel}, D.~J., {Strauss}, M.~A., {et~al.} 2005, \aj, 129,
  2562

\bibitem[{{Braspenning} {et~al.}(2023){Braspenning}, {Schaye}, {Schaller},
  {McCarthy}, {Kay}, {Helly}, {Kugel}, {Elbers}, {Frenk}, {Kwan}, {Salcido},
  {van Daalen}, \& {Vandenbroucke}}]{Braspenning2023}
{Braspenning}, J., {Schaye}, J., {Schaller}, M., {et~al.} 2023, arXiv e-prints,
  arXiv:2312.08277

\bibitem[{{Brinchmann} {et~al.}(2004){Brinchmann}, {Charlot}, {White},
  {Tremonti}, {Kauffmann}, {Heckman}, \& {Brinkmann}}]{Brinchmann2004}
{Brinchmann}, J., {Charlot}, S., {White}, S.~D.~M., {et~al.} 2004, \mnras, 351,
  1151

\bibitem[{{Bulbul} {et~al.}(2019){Bulbul}, {Chiu}, {Mohr}, {McDonald},
  {Benson}, {Bautz}, {Bayliss}, {Bleem}, {Brodwin}, {Bocquet}, {Capasso},
  {Dietrich}, {Forman}, {Hlavacek-Larrondo}, {Holzapfel}, {Khullar}, {Klein},
  {Kraft}, {Miller}, {Reichardt}, {Saro}, {Sharon}, {Stalder}, {Schrabback}, \&
  {Stanford}}]{Bulbul2019}
{Bulbul}, E., {Chiu}, I.~N., {Mohr}, J.~J., {et~al.} 2019, \apj, 871, 50

\bibitem[{{Chadayammuri} {et~al.}(2022){Chadayammuri}, {Bogd{\'a}n},
  {Oppenheimer}, {Kraft}, {Forman}, \& {Jones}}]{Chada2022}
{Chadayammuri}, U., {Bogd{\'a}n}, {\'A}., {Oppenheimer}, B.~D., {et~al.} 2022,
  \apjl, 936, L15

\bibitem[{{Chen} {et~al.}(2012){Chen}, {Kauffmann}, {Tremonti}, {White},
  {Heckman}, {Kova{\v{c}}}, {Bundy}, {Chisholm}, {Maraston}, {Schneider},
  {Bolton}, {Weaver}, \& {Brinkmann}}]{Chen2012}
{Chen}, Y.-M., {Kauffmann}, G., {Tremonti}, C.~A., {et~al.} 2012, \mnras, 421,
  314

\bibitem[{{Chiu} {et~al.}(2022){Chiu}, {Ghirardini}, {Liu}, {Grandis},
  {Bulbul}, {Bahar}, {Comparat}, {Bocquet}, {Clerc}, {Klein}, {Liu}, {Li},
  {Miyatake}, {Mohr}, {More}, {Oguri}, {Okabe}, {Pacaud}, {Ramos-Ceja},
  {Reiprich}, {Schrabback}, \& {Umetsu}}]{Chiu2022}
{Chiu}, I.~N., {Ghirardini}, V., {Liu}, A., {et~al.} 2022, \aap, 661, A11

\bibitem[{{Choi} {et~al.}(2015){Choi}, {Ostriker}, {Naab}, {Oser}, \&
  {Moster}}]{Choi2015}
{Choi}, E., {Ostriker}, J.~P., {Naab}, T., {Oser}, L., \& {Moster}, B.~P. 2015,
  \mnras, 449, 4105

\bibitem[{{Comparat} {et~al.}(2023){Comparat}, {Luo}, {Merloni}, {More},
  {Salvato}, {Krumpe}, {Miyaji}, {Brandt}, {Georgakakis}, {Akiyama}, {Buchner},
  {Dwelly}, {Kawaguchi}, {Liu}, {Nagao}, {Nandra}, {Silverman}, {Toba},
  {Anderson}, \& {Kollmeier}}]{Comparat2023}
{Comparat}, J., {Luo}, W., {Merloni}, A., {et~al.} 2023, \aap, 673, A122

\bibitem[{{Comparat} {et~al.}(2019){Comparat}, {Merloni}, {Salvato}, {Nandra},
  {Boller}, {Georgakakis}, {Finoguenov}, {Dwelly}, {Buchner}, {Del Moro},
  {Clerc}, {Wang}, {Zhao}, {Prada}, {Yepes}, {Brusa}, {Krumpe}, \&
  {Liu}}]{Comparat2019}
{Comparat}, J., {Merloni}, A., {Salvato}, M., {et~al.} 2019, \mnras, 487, 2005

\bibitem[{{Comparat} {et~al.}(2022){Comparat}, {Truong}, {Merloni},
  {Pillepich}, {Ponti}, {Driver}, {Bellstedt}, {Liske}, {Aird}, {Br{\"u}ggen},
  {Bulbul}, {Davies}, {Villalba}, {Georgakakis}, {Haberl}, {Liu}, {Maitra},
  {Nandra}, {Popesso}, {Predehl}, {Robotham}, {Salvato}, {Thorne}, \&
  {Zhang}}]{Comparat2022}
{Comparat}, J., {Truong}, N., {Merloni}, A., {et~al.} 2022, \aap, 666, A156

\bibitem[{{Coupon} {et~al.}(2015){Coupon}, {Arnouts}, {van Waerbeke},
  {Moutard}, {Ilbert}, {van Uitert}, {Erben}, {Garilli}, {Guzzo}, {Heymans},
  {Hildebrandt}, {Hoekstra}, {Kilbinger}, {Kitching}, {Mellier}, {Miller},
  {Scodeggio}, {Bonnett}, {Branchini}, {Davidzon}, {De Lucia}, {Fritz}, {Fu},
  {Hudelot}, {Hudson}, {Kuijken}, {Leauthaud}, {Le F{\`e}vre}, {McCracken},
  {Moscardini}, {Rowe}, {Schrabback}, {Semboloni}, \& {Velander}}]{Coupon2015}
{Coupon}, J., {Arnouts}, S., {van Waerbeke}, L., {et~al.} 2015, \mnras, 449,
  1352

\bibitem[{{Crain} {et~al.}(2015){Crain}, {Schaye}, {Bower}, {Furlong},
  {Schaller}, {Theuns}, {Dalla Vecchia}, {Frenk}, {McCarthy}, {Helly},
  {Jenkins}, {Rosas-Guevara}, {White}, \& {Trayford}}]{Crain2015}
{Crain}, R.~A., {Schaye}, J., {Bower}, R.~G., {et~al.} 2015, \mnras, 450, 1937

\bibitem[{{Das} {et~al.}(2023){Das}, {Chiang}, \&
  {Mathur}}]{DasChiangMathur_2023ApJ...951..125D}
{Das}, S., {Chiang}, Y.-K., \& {Mathur}, S. 2023, \apj, 951, 125

\bibitem[{{Dav{\'e}} {et~al.}(2019){Dav{\'e}}, {Angl{\'e}s-Alc{\'a}zar},
  {Narayanan}, {Li}, {Rafieferantsoa}, \& {Appleby}}]{Dave2019}
{Dav{\'e}}, R., {Angl{\'e}s-Alc{\'a}zar}, D., {Narayanan}, D., {et~al.} 2019,
  \mnras, 486, 2827

\bibitem[{{Davies} {et~al.}(2019){Davies}, {Crain}, {McCarthy}, {Oppenheimer},
  {Schaye}, {Schaller}, \& {McAlpine}}]{Davies2019}
{Davies}, J.~J., {Crain}, R.~A., {McCarthy}, I.~G., {et~al.} 2019, \mnras, 485,
  3783

\bibitem[{{de Jong} {et~al.}(2019){de Jong}, {Agertz}, {Berbel}, {Aird},
  {Alexander}, {Amarsi}, {Anders}, {Andrae}, {Ansarinejad}, {Ansorge}, \&
  et~al.}]{deJongAgertzBerbel_2019Msngr.175....3D}
{de Jong}, R.~S., {Agertz}, O., {Berbel}, A.~A., {et~al.} 2019, The Messenger,
  175, 3

\bibitem[{{Dey} {et~al.}(2019){Dey}, {Schlegel}, {Lang}, {Blum}, {Burleigh},
  {Fan}, {Findlay}, {Finkbeiner}, {Herrera}, {Juneau}, {Landriau}, {Levi},
  {McGreer}, {Meisner}, {Myers}, {Moustakas}, {Nugent}, {Patej}, {Schlafly},
  {Walker}, {Valdes}, {Weaver}, {Y{\`e}che}, {Zou}, {Zhou}, {Abareshi},
  {Abbott}, {Abolfathi}, {Aguilera}, {Alam}, {Allen}, {Alvarez}, {Annis},
  {Ansarinejad}, {Aubert}, {Beechert}, {Bell}, {BenZvi}, {Beutler}, {Bielby},
  {Bolton}, {Brice{\~n}o}, {Buckley-Geer}, {Butler}, {Calamida}, {Carlberg},
  {Carter}, {Casas}, {Castander}, {Choi}, {Comparat}, {Cukanovaite}, {Delubac},
  {DeVries}, {Dey}, {Dhungana}, {Dickinson}, {Ding}, {Donaldson}, {Duan},
  {Duckworth}, {Eftekharzadeh}, {Eisenstein}, {Etourneau}, {Fagrelius},
  {Farihi}, {Fitzpatrick}, {Font-Ribera}, {Fulmer}, {G{\"a}nsicke},
  {Gaztanaga}, {George}, {Gerdes}, {Gontcho}, {Gorgoni}, {Green}, {Guy},
  {Harmer}, {Hernandez}, {Honscheid}, {Huang}, {James}, {Jannuzi}, {Jiang},
  {Joyce}, {Karcher}, {Karkar}, {Kehoe}, {Kneib}, {Kueter-Young}, {Lan},
  {Lauer}, {Le Guillou}, {Le Van Suu}, {Lee}, {Lesser}, {Perreault Levasseur},
  {Li}, {Mann}, {Marshall}, {Mart{\'\i}nez-V{\'a}zquez}, {Martini}, {du Mas des
  Bourboux}, {McManus}, {Meier}, {M{\'e}nard}, {Metcalfe},
  {Mu{\~n}oz-Guti{\'e}rrez}, {Najita}, {Napier}, {Narayan}, {Newman}, {Nie},
  {Nord}, {Norman}, {Olsen}, {Paat}, {Palanque-Delabrouille}, {Peng},
  {Poppett}, {Poremba}, {Prakash}, {Rabinowitz}, {Raichoor}, {Rezaie},
  {Robertson}, {Roe}, {Ross}, {Ross}, {Rudnick}, {Safonova}, {Saha},
  {S{\'a}nchez}, {Savary}, {Schweiker}, {Scott}, {Seo}, {Shan}, {Silva},
  {Slepian}, {Soto}, {Sprayberry}, {Staten}, {Stillman}, {Stupak}, {Summers},
  {Sien Tie}, {Tirado}, {Vargas-Maga{\~n}a}, {Vivas}, {Wechsler}, {Williams},
  {Yang}, {Yang}, {Yapici}, {Zaritsky}, {Zenteno}, {Zhang}, {Zhang}, {Zhou}, \&
  {Zhou}}]{Dey2019}
{Dey}, A., {Schlegel}, D.~J., {Lang}, D., {et~al.} 2019, \aj, 157, 168

\bibitem[{{Eckert} {et~al.}(2021){Eckert}, {Gaspari}, {Gastaldello}, {Le Brun},
  \& {O'Sullivan}}]{EckertGaspariGastaldello_2021Univ....7..142E}
{Eckert}, D., {Gaspari}, M., {Gastaldello}, F., {Le Brun}, A. M.~C., \&
  {O'Sullivan}, E. 2021, Universe, 7, 142

\bibitem[{{Eckmiller} {et~al.}(2011){Eckmiller}, {Hudson}, \&
  {Reiprich}}]{Eckmiller2011}
{Eckmiller}, H.~J., {Hudson}, D.~S., \& {Reiprich}, T.~H. 2011, \aap, 535, A105

\bibitem[{{Fabian}(2012)}]{Fabian2012}
{Fabian}, A.~C. 2012, \araa, 50, 455

\bibitem[{{Faucher-Gigu{\`e}re} \& {Oh}(2023)}]{Faucher2023}
{Faucher-Gigu{\`e}re}, C.-A. \& {Oh}, S.~P. 2023, \araa, 61, 131

\bibitem[{{Finoguenov} {et~al.}(2019){Finoguenov}, {Merloni}, {Comparat},
  {Nandra}, {Salvato}, {Tempel}, {Raichoor}, {Richard}, {Kneib}, {Pillepich},
  \& et~al.}]{FinoguenovMerloniComparat_2019Msngr.175...39F}
{Finoguenov}, A., {Merloni}, A., {Comparat}, J., {et~al.} 2019, The Messenger,
  175, 39

\bibitem[{{Forbes} {et~al.}(2017){Forbes}, {Alabi}, {Romanowsky}, {Kim},
  {Brodie}, \& {Fabbiano}}]{Forbes2017}
{Forbes}, D.~A., {Alabi}, A., {Romanowsky}, A.~J., {et~al.} 2017, \mnras, 464,
  L26

\bibitem[{{Goulding} {et~al.}(2016){Goulding}, {Greene}, {Ma}, {Veale},
  {Bogdan}, {Nyland}, {Blakeslee}, {McConnell}, \& {Thomas}}]{Goulding2016}
{Goulding}, A.~D., {Greene}, J.~E., {Ma}, C.-P., {et~al.} 2016, \apj, 826, 167

\bibitem[{{Hahn} {et~al.}(2023){Hahn}, {Wilson}, {Ruiz-Macias}, {Cole},
  {Weinberg}, {Moustakas}, {Kremin}, {Tinker}, {Smith}, {Wechsler}, \&
  et~al.}]{HahnWilsonRuiz-Macias_2023AJ....165..253H}
{Hahn}, C., {Wilson}, M.~J., {Ruiz-Macias}, O., {et~al.} 2023, \aj, 165, 253

\bibitem[{{Hogg} {et~al.}(2010){Hogg}, {Bovy}, \& {Lang}}]{emcee2010}
{Hogg}, D.~W., {Bovy}, J., \& {Lang}, D. 2010, arXiv e-prints, arXiv:1008.4686

\bibitem[{{Ishiyama} {et~al.}(2021){Ishiyama}, {Prada}, {Klypin}, {Sinha},
  {Metcalf}, {Jullo}, {Altieri}, {Cora}, {Croton}, {de la Torre},
  {Mill{\'a}n-Calero}, {Oogi}, {Ruedas}, \&
  {Vega-Mart{\'\i}nez}}]{Ishiyama2021}
{Ishiyama}, T., {Prada}, F., {Klypin}, A.~A., {et~al.} 2021, \mnras, 506, 4210

\bibitem[{{Iyer} {et~al.}(2018){Iyer}, {Gawiser}, {Dav{\'e}}, {Davis},
  {Finkelstein}, {Kodra}, {Koekemoer}, {Kurczynski}, {Newman}, {Pacifici}, \&
  {Somerville}}]{Iyer2018}
{Iyer}, K., {Gawiser}, E., {Dav{\'e}}, R., {et~al.} 2018, \apj, 866, 120

\bibitem[{{Kaiser}(1986)}]{Kaiser1986}
{Kaiser}, N. 1986, \mnras, 222, 323

\bibitem[{{Kim} \& {Fabbiano}(2013)}]{Kim2013}
{Kim}, D.-W. \& {Fabbiano}, G. 2013, \apj, 776, 116

\bibitem[{{Kim} \& {Fabbiano}(2015)}]{Kim2015}
{Kim}, D.-W. \& {Fabbiano}, G. 2015, \apj, 812, 127

\bibitem[{{Kravtsov} \& {Borgani}(2012)}]{Kravtsov2012}
{Kravtsov}, A.~V. \& {Borgani}, S. 2012, \araa, 50, 353

\bibitem[{{Kyritsis} {et~al.}(2024){Kyritsis}, {Zezas}, {Haberl}, {Weber},
  {Basu-Zych}, {Vulic}, {Maitra}, {H{\"a}mmerich}, {Wilms}, {Sasaki},
  {Hornschemeier}, {Ptak}, {Merloni}, \& {Comparat}}]{Kyritsis2024}
{Kyritsis}, E., {Zezas}, A., {Haberl}, F., {et~al.} 2024, arXiv e-prints,
  arXiv:2402.12367

\bibitem[{{Le Brun} {et~al.}(2014){Le Brun}, {McCarthy}, {Schaye}, \&
  {Ponman}}]{LeBrun2014}
{Le Brun}, A. M.~C., {McCarthy}, I.~G., {Schaye}, J., \& {Ponman}, T.~J. 2014,
  \mnras, 441, 1270

\bibitem[{{Leauthaud} {et~al.}(2012{\natexlab{a}}){Leauthaud}, {George},
  {Behroozi}, {Bundy}, {Tinker}, {Wechsler}, {Conroy}, {Finoguenov}, \&
  {Tanaka}}]{Leauthaud2012b}
{Leauthaud}, A., {George}, M.~R., {Behroozi}, P.~S., {et~al.}
  2012{\natexlab{a}}, \apj, 746, 95

\bibitem[{{Leauthaud} {et~al.}(2012{\natexlab{b}}){Leauthaud}, {Tinker},
  {Bundy}, {Behroozi}, {Massey}, {Rhodes}, {George}, {Kneib}, {Benson},
  {Wechsler}, {Busha}, {Capak}, {Cort{\^e}s}, {Ilbert}, {Koekemoer}, {Le
  F{\`e}vre}, {Lilly}, {McCracken}, {Salvato}, {Schrabback}, {Scoville},
  {Smith}, \& {Taylor}}]{Leauthaud2012a}
{Leauthaud}, A., {Tinker}, J., {Bundy}, K., {et~al.} 2012{\natexlab{b}}, \apj,
  744, 159

\bibitem[{{Lehmer} {et~al.}(2019){Lehmer}, {Eufrasio}, {Tzanavaris},
  {Basu-Zych}, {Fragos}, {Prestwich}, {Yukita}, {Zezas}, {Hornschemeier}, \&
  {Ptak}}]{Lehmer2019}
{Lehmer}, B.~D., {Eufrasio}, R.~T., {Tzanavaris}, P., {et~al.} 2019, \apjs,
  243, 3

\bibitem[{{Li} \& {Wang}(2013)}]{Li2013a}
{Li}, J.-T. \& {Wang}, Q.~D. 2013, \mnras, 435, 3071

\bibitem[{{Liu} {et~al.}(2022){Liu}, {Bulbul}, {Ghirardini}, {Liu}, {Klein},
  {Clerc}, {{\"O}zsoy}, {Ramos-Ceja}, {Pacaud}, {Comparat}, \&
  et~al.}]{Liu2022}
{Liu}, A., {Bulbul}, E., {Ghirardini}, V., {et~al.} 2022, \aap, 661, A2

\bibitem[{{Lovisari} {et~al.}(2021){Lovisari}, {Ettori}, {Gaspari}, \&
  {Giles}}]{Lovisari2021}
{Lovisari}, L., {Ettori}, S., {Gaspari}, M., \& {Giles}, P.~A. 2021, Universe,
  7, 139

\bibitem[{{Lovisari} \& {Maughan}(2022)}]{Lovisari2022}
{Lovisari}, L. \& {Maughan}, B.~J. 2022, in Handbook of X-ray and Gamma-ray
  Astrophysics, 65

\bibitem[{{Lovisari} {et~al.}(2015){Lovisari}, {Reiprich}, \&
  {Schellenberger}}]{Lovisari2015}
{Lovisari}, L., {Reiprich}, T.~H., \& {Schellenberger}, G. 2015, \aap, 573,
  A118

\bibitem[{{Lovisari} {et~al.}(2020){Lovisari}, {Schellenberger}, {Sereno},
  {Ettori}, {Pratt}, {Forman}, {Jones}, {Andrade-Santos}, {Randall}, \&
  {Kraft}}]{Lovisari2020}
{Lovisari}, L., {Schellenberger}, G., {Sereno}, M., {et~al.} 2020, \apj, 892,
  102

\bibitem[{{Mandelbaum} {et~al.}(2016){Mandelbaum}, {Wang}, {Zu}, {White},
  {Henriques}, \& {More}}]{MandelbaumWangZu_2016MNRAS.457.3200M}
{Mandelbaum}, R., {Wang}, W., {Zu}, Y., {et~al.} 2016, \mnras, 457, 3200

\bibitem[{{Mantz} {et~al.}(2016){Mantz}, {Allen}, {Morris}, \&
  {Schmidt}}]{Mantz2016}
{Mantz}, A.~B., {Allen}, S.~W., {Morris}, R.~G., \& {Schmidt}, R.~W. 2016,
  \mnras, 456, 4020

\bibitem[{{Marinacci} {et~al.}(2018){Marinacci}, {Vogelsberger}, {Pakmor},
  {Torrey}, {Springel}, {Hernquist}, {Nelson}, {Weinberger}, {Pillepich},
  {Naiman}, \& {Genel}}]{Marinacci2018}
{Marinacci}, F., {Vogelsberger}, M., {Pakmor}, R., {et~al.} 2018, \mnras, 480,
  5113

\bibitem[{{McAlpine} {et~al.}(2016){McAlpine}, {Helly}, {Schaller}, {Trayford},
  {Qu}, {Furlong}, {Bower}, {Crain}, {Schaye}, {Theuns}, {Dalla Vecchia},
  {Frenk}, {McCarthy}, {Jenkins}, {Rosas-Guevara}, {White}, {Baes}, {Camps}, \&
  {Lemson}}]{McAlpine2016}
{McAlpine}, S., {Helly}, J.~C., {Schaller}, M., {et~al.} 2016, Astronomy and
  Computing, 15, 72

\bibitem[{{McIntosh}(2016)}]{McIntoshjackknife2016}
{McIntosh}, A. 2016, arXiv e-prints, arXiv:1606.00497

\bibitem[{{Merloni} {et~al.}(2024){Merloni}, {Lamer}, {Liu}, {Ramos-Ceja},
  {Brunner}, {Bulbul}, {Liu}, {Ghirardini}, {Liu}, {Salvato}, {Sanders},
  {Wilms}, {Dwelly}, {Dauser}, {K{\"o}nig}, {Ramos-Ceja}, {Garrel}, \&
  {Reiprich}}]{Merloni2024}
{Merloni}, A., {Lamer}, G., {Liu}, T., {et~al.} 2024, \aap, 682, A78

\bibitem[{{Merloni} {et~al.}(2012){Merloni}, {Predehl}, {Becker},
  {B{\"o}hringer}, {Boller}, {Brunner}, {Brusa}, {Dennerl}, {Freyberg},
  {Friedrich}, {Georgakakis}, {Haberl}, {Hasinger}, {Meidinger}, {Mohr},
  {Nandra}, {Rau}, {Reiprich}, {Robrade}, {Salvato}, {Santangelo}, {Sasaki},
  {Schwope}, {Wilms}, \& {German eROSITA Consortium}}]{Merloni2012}
{Merloni}, A., {Predehl}, P., {Becker}, W., {et~al.} 2012, arXiv e-prints,
  arXiv:1209.3114

\bibitem[{{Mowla} {et~al.}(2019{\natexlab{a}}){Mowla}, {van der Wel}, {van
  Dokkum}, \& {Miller}}]{Mowla2019a}
{Mowla}, L., {van der Wel}, A., {van Dokkum}, P., \& {Miller}, T.~B.
  2019{\natexlab{a}}, \apjl, 872, L13

\bibitem[{{Mowla} {et~al.}(2019{\natexlab{b}}){Mowla}, {van Dokkum}, {Brammer},
  {Momcheva}, {van der Wel}, {Whitaker}, {Nelson}, {Bezanson}, {Muzzin},
  {Franx}, {MacKenty}, {Leja}, {Kriek}, \& {Marchesini}}]{Mowla2019b}
{Mowla}, L.~A., {van Dokkum}, P., {Brammer}, G.~B., {et~al.}
  2019{\natexlab{b}}, \apj, 880, 57

\bibitem[{{Muldrew} {et~al.}(2012){Muldrew}, {Croton}, {Skibba}, {Pearce},
  {Ann}, {Baldry}, {Brough}, {Choi}, {Conselice}, {Cowan}, {Gallazzi}, {Gray},
  {Gr{\"u}tzbauch}, {Li}, {Park}, {Pilipenko}, {Podgorzec}, {Robotham},
  {Wilman}, {Yang}, {Zhang}, \& {Zibetti}}]{Muldrew2012}
{Muldrew}, S.~I., {Croton}, D.~J., {Skibba}, R.~A., {et~al.} 2012, \mnras, 419,
  2670

\bibitem[{{Naiman} {et~al.}(2018){Naiman}, {Pillepich}, {Springel},
  {Ramirez-Ruiz}, {Torrey}, {Vogelsberger}, {Pakmor}, {Nelson}, {Marinacci},
  {Hernquist}, {Weinberger}, \& {Genel}}]{Naiman2018}
{Naiman}, J.~P., {Pillepich}, A., {Springel}, V., {et~al.} 2018, \mnras, 477,
  1206

\bibitem[{{Nelson} {et~al.}(2019{\natexlab{a}}){Nelson}, {Pillepich},
  {Springel}, {Pakmor}, {Weinberger}, {Genel}, {Torrey}, {Vogelsberger},
  {Marinacci}, \& {Hernquist}}]{Nelson2019b}
{Nelson}, D., {Pillepich}, A., {Springel}, V., {et~al.} 2019{\natexlab{a}},
  \mnras, 490, 3234

\bibitem[{{Nelson} {et~al.}(2019{\natexlab{b}}){Nelson}, {Springel},
  {Pillepich}, {Rodriguez-Gomez}, {Torrey}, {Genel}, {Vogelsberger}, {Pakmor},
  {Marinacci}, {Weinberger}, {Kelley}, {Lovell}, {Diemer}, \&
  {Hernquist}}]{Nelson2019a}
{Nelson}, D., {Springel}, V., {Pillepich}, A., {et~al.} 2019{\natexlab{b}},
  Computational Astrophysics and Cosmology, 6, 2

\bibitem[{{Oppenheimer} {et~al.}(2021){Oppenheimer}, {Babul}, {Bah{\'e}},
  {Butsky}, \& {McCarthy}}]{OppenheimerBabulBahe_2021Univ....7..209O}
{Oppenheimer}, B.~D., {Babul}, A., {Bah{\'e}}, Y., {Butsky}, I.~S., \&
  {McCarthy}, I.~G. 2021, Universe, 7, 209

\bibitem[{{Pandey} {et~al.}(2022){Pandey}, {Gatti}, {Baxter}, {Hill}, {Fang},
  {Doux}, {Giannini}, {Raveri}, {DeRose}, {Huang}, \&
  et~al.}]{PandeyGattiBaxter_2022PhRvD.105l3526P}
{Pandey}, S., {Gatti}, M., {Baxter}, E., {et~al.} 2022, \prd, 105, 123526

\bibitem[{{Pillepich} {et~al.}(2018){Pillepich}, {Nelson}, {Hernquist},
  {Springel}, {Pakmor}, {Torrey}, {Weinberger}, {Genel}, {Naiman}, {Marinacci},
  \& {Vogelsberger}}]{Pillepich2018}
{Pillepich}, A., {Nelson}, D., {Hernquist}, L., {et~al.} 2018, \mnras, 475, 648

\bibitem[{{Planck Collaboration} {et~al.}(2013){Planck Collaboration}, {Ade},
  {Aghanim}, {Arnaud}, {Ashdown}, {Atrio-Barandela}, {Aumont}, {Baccigalupi},
  {Balbi}, {Banday}, \&
  et~al.}]{PlanckCollaborationAdeAghanim_2013A&A...557A..52P}
{Planck Collaboration}, {Ade}, P.~A.~R., {Aghanim}, N., {et~al.} 2013, \aap,
  557, A52

\bibitem[{{Planck Collaboration} {et~al.}(2020){Planck Collaboration},
  {Aghanim}, {Akrami}, {Ashdown}, {Aumont}, {Baccigalupi}, {Ballardini},
  {Banday}, {Barreiro}, {Bartolo}, {Basak}, {Battye}, {Benabed}, {Bernard},
  {Bersanelli}, {Bielewicz}, {Bock}, {Bond}, {Borrill}, {Bouchet}, {Boulanger},
  {Bucher}, {Burigana}, {Butler}, {Calabrese}, {Cardoso}, {Carron},
  {Challinor}, {Chiang}, {Chluba}, {Colombo}, {Combet}, {Contreras}, {Crill},
  {Cuttaia}, {de Bernardis}, {de Zotti}, {Delabrouille}, {Delouis}, {Di
  Valentino}, {Diego}, {Dor{\'e}}, {Douspis}, {Ducout}, {Dupac}, {Dusini},
  {Efstathiou}, {Elsner}, {En{\ss}lin}, {Eriksen}, {Fantaye}, {Farhang},
  {Fergusson}, {Fernandez-Cobos}, {Finelli}, {Forastieri}, {Frailis},
  {Fraisse}, {Franceschi}, {Frolov}, {Galeotta}, {Galli}, {Ganga},
  {G{\'e}nova-Santos}, {Gerbino}, {Ghosh}, {Gonz{\'a}lez-Nuevo}, {G{\'o}rski},
  {Gratton}, {Gruppuso}, {Gudmundsson}, {Hamann}, {Handley}, {Hansen},
  {Herranz}, {Hildebrandt}, {Hivon}, {Huang}, {Jaffe}, {Jones}, {Karakci},
  {Keih{\"a}nen}, {Keskitalo}, {Kiiveri}, {Kim}, {Kisner}, {Knox},
  {Krachmalnicoff}, {Kunz}, {Kurki-Suonio}, {Lagache}, {Lamarre}, {Lasenby},
  {Lattanzi}, {Lawrence}, {Le Jeune}, {Lemos}, {Lesgourgues}, {Levrier},
  {Lewis}, {Liguori}, {Lilje}, {Lilley}, {Lindholm}, {L{\'o}pez-Caniego},
  {Lubin}, {Ma}, {Mac{\'\i}as-P{\'e}rez}, {Maggio}, {Maino}, {Mandolesi},
  {Mangilli}, {Marcos-Caballero}, {Maris}, {Martin}, {Martinelli},
  {Mart{\'\i}nez-Gonz{\'a}lez}, {Matarrese}, {Mauri}, {McEwen}, {Meinhold},
  {Melchiorri}, {Mennella}, {Migliaccio}, {Millea}, {Mitra},
  {Miville-Desch{\^e}nes}, {Molinari}, {Montier}, {Morgante}, {Moss}, {Natoli},
  {N{\o}rgaard-Nielsen}, {Pagano}, {Paoletti}, {Partridge}, {Patanchon},
  {Peiris}, {Perrotta}, {Pettorino}, {Piacentini}, {Polastri}, {Polenta},
  {Puget}, {Rachen}, {Reinecke}, {Remazeilles}, {Renzi}, {Rocha}, {Rosset},
  {Roudier}, {Rubi{\~n}o-Mart{\'\i}n}, {Ruiz-Granados}, {Salvati}, {Sandri},
  {Savelainen}, {Scott}, {Shellard}, {Sirignano}, {Sirri}, {Spencer},
  {Sunyaev}, {Suur-Uski}, {Tauber}, {Tavagnacco}, {Tenti}, {Toffolatti},
  {Tomasi}, {Trombetti}, {Valenziano}, {Valiviita}, {Van Tent}, {Vibert},
  {Vielva}, {Villa}, {Vittorio}, {Wandelt}, {Wehus}, {White}, {White},
  {Zacchei}, \& {Zonca}}]{Planck2020}
{Planck Collaboration}, {Aghanim}, N., {Akrami}, Y., {et~al.} 2020, \aap, 641,
  A6

\bibitem[{{Pop} {et~al.}(2022){Pop}, {Hernquist}, {Nagai}, {Kannan},
  {Weinberger}, {Springel}, {Vogelsberger}, {Nelson}, {Pakmor}, {Pillepich}, \&
  {Torrey}}]{Pop2022}
{Pop}, A.-R., {Hernquist}, L., {Nagai}, D., {et~al.} 2022, arXiv e-prints,
  arXiv:2205.11528

\bibitem[{{Popesso} {et~al.}(2024){Popesso}, {Biviano}, {Bulbul}, {Merloni},
  {Comparat}, {Clerc}, {Igo}, {Liu}, {Driver}, {Salvato}, {Brusa}, {Bahar},
  {Malavasi}, {Ghirardini}, {Robotham}, {Liske}, \& {Grandis}}]{Popesso2024}
{Popesso}, P., {Biviano}, A., {Bulbul}, E., {et~al.} 2024, \mnras, 527, 895

\bibitem[{{Predehl} {et~al.}(2021){Predehl}, {Andritschke}, {Arefiev},
  {Babyshkin}, {Batanov}, {Becker}, {B{\"o}hringer}, {Bogomolov}, {Boller},
  {Borm}, {Bornemann}, {Br{\"a}uninger}, {Br{\"u}ggen}, {Brunner}, {Brusa},
  {Bulbul}, {Buntov}, {Burwitz}, {Burkert}, {Clerc}, {Churazov}, {Coutinho},
  {Dauser}, {Dennerl}, {Doroshenko}, {Eder}, {Emberger}, {Eraerds},
  {Finoguenov}, {Freyberg}, {Friedrich}, {Friedrich}, {F{\"u}rmetz},
  {Georgakakis}, {Gilfanov}, {Granato}, {Grossberger}, {Gueguen}, {Gureev},
  {Haberl}, {H{\"a}lker}, {Hartner}, {Hasinger}, {Huber}, {Ji}, {Kienlin},
  {Kink}, {Korotkov}, {Kreykenbohm}, {Lamer}, {Lomakin}, {Lapshov}, {Liu},
  {Maitra}, {Meidinger}, {Menz}, {Merloni}, {Mernik}, {Mican}, {Mohr},
  {M{\"u}ller}, {Nandra}, {Nazarov}, {Pacaud}, {Pavlinsky}, {Perinati},
  {Pfeffermann}, {Pietschner}, {Ramos-Ceja}, {Rau}, {Reiffers}, {Reiprich},
  {Robrade}, {Salvato}, {Sanders}, {Santangelo}, {Sasaki}, {Scheuerle},
  {Schmid}, {Schmitt}, {Schwope}, {Shirshakov}, {Steinmetz}, {Stewart},
  {Str{\"u}der}, {Sunyaev}, {Tenzer}, {Tiedemann}, {Tr{\"u}mper}, {Voron},
  {Weber}, {Wilms}, \& {Yaroshenko}}]{Predehl2021}
{Predehl}, P., {Andritschke}, R., {Arefiev}, V., {et~al.} 2021, \aap, 647, A1

\bibitem[{{Schaan} {et~al.}(2021){Schaan}, {Ferraro}, {Amodeo}, {Battaglia},
  {Aiola}, {Austermann}, {Beall}, {Bean}, {Becker}, {Bond}, \&
  et~al.}]{SchaanFerraroAmodeo_2021PhRvD.103f3513S}
{Schaan}, E., {Ferraro}, S., {Amodeo}, S., {et~al.} 2021, \prd, 103, 063513

\bibitem[{{Schaller} {et~al.}(2015){Schaller}, {Dalla Vecchia}, {Schaye},
  {Bower}, {Theuns}, {Crain}, {Furlong}, \& {McCarthy}}]{Schaller2015}
{Schaller}, M., {Dalla Vecchia}, C., {Schaye}, J., {et~al.} 2015, \mnras, 454,
  2277

\bibitem[{{Schaye} {et~al.}(2015){Schaye}, {Crain}, {Bower}, {Furlong},
  {Schaller}, {Theuns}, {Dalla Vecchia}, {Frenk}, {McCarthy}, {Helly},
  {Jenkins}, {Rosas-Guevara}, {White}, {Baes}, {Booth}, {Camps}, {Navarro},
  {Qu}, {Rahmati}, {Sawala}, {Thomas}, \& {Trayford}}]{Schaye2015}
{Schaye}, J., {Crain}, R.~A., {Bower}, R.~G., {et~al.} 2015, \mnras, 446, 521

\bibitem[{{Schaye} {et~al.}(2023){Schaye}, {Kugel}, {Schaller}, {Helly},
  {Braspenning}, {Elbers}, {McCarthy}, {van Daalen}, {Vandenbroucke}, {Frenk},
  {Kwan}, {Salcido}, {Bah{\'e}}, {Borrow}, {Chaikin}, {Hahn}, {Hu{\v{s}}ko},
  {Jenkins}, {Lacey}, \& {Nobels}}]{Schaye2023}
{Schaye}, J., {Kugel}, R., {Schaller}, M., {et~al.} 2023, \mnras, 526, 4978

\bibitem[{{Schellenberger} \& {Reiprich}(2017)}]{Schellenberger2017}
{Schellenberger}, G. \& {Reiprich}, T.~H. 2017, \mnras, 469, 3738

\bibitem[{{Smith} {et~al.}(2001){Smith}, {Brickhouse}, {Liedahl}, \&
  {Raymond}}]{Smith2001}
{Smith}, R.~K., {Brickhouse}, N.~S., {Liedahl}, D.~A., \& {Raymond}, J.~C.
  2001, \apjl, 556, L91

\bibitem[{{Springel} {et~al.}(2018){Springel}, {Pakmor}, {Pillepich},
  {Weinberger}, {Nelson}, {Hernquist}, {Vogelsberger}, {Genel}, {Torrey},
  {Marinacci}, \& {Naiman}}]{Springel2018}
{Springel}, V., {Pakmor}, R., {Pillepich}, A., {et~al.} 2018, \mnras, 475, 676

\bibitem[{{Stern} {et~al.}(2021){Stern}, {Faucher-Gigu{\`e}re}, {Fielding},
  {Quataert}, {Hafen}, {Gurvich}, {Ma}, {Byrne}, {El-Badry},
  {Angl{\'e}s-Alc{\'a}zar}, {Chan}, {Feldmann}, {Kere{\v{s}}}, {Wetzel},
  {Murray}, \& {Hopkins}}]{Stern2021}
{Stern}, J., {Faucher-Gigu{\`e}re}, C.-A., {Fielding}, D., {et~al.} 2021, \apj,
  911, 88

\bibitem[{{Strauss} {et~al.}(2002){Strauss}, {Weinberg}, {Lupton}, {Narayanan},
  {Annis}, {Bernardi}, {Blanton}, {Burles}, {Connolly}, {Dalcanton}, {Doi},
  {Eisenstein}, {Frieman}, {Fukugita}, {Gunn}, {Ivezi{\'c}}, {Kent}, {Kim},
  {Knapp}, {Kron}, {Munn}, {Newberg}, {Nichol}, {Okamura}, {Quinn}, {Richmond},
  {Schlegel}, {Shimasaku}, {SubbaRao}, {Szalay}, {Vanden Berk}, {Vogeley},
  {Yanny}, {Yasuda}, {York}, \& {Zehavi}}]{Strauss2002}
{Strauss}, M.~A., {Weinberg}, D.~H., {Lupton}, R.~H., {et~al.} 2002, \aj, 124,
  1810

\bibitem[{{Sunyaev} {et~al.}(2021){Sunyaev}, {Arefiev}, {Babyshkin},
  {Bogomolov}, {Borisov}, {Buntov}, {Brunner}, {Burenin}, {Churazov},
  {Coutinho}, \& et~al.}]{SunyaevArefievBabyshkin_2021A&A...656A.132S}
{Sunyaev}, R., {Arefiev}, V., {Babyshkin}, V., {et~al.} 2021, \aap, 656, A132

\bibitem[{{Taylor} {et~al.}(2011){Taylor}, {Hopkins}, {Baldry}, {Brown},
  {Driver}, {Kelvin}, {Hill}, {Robotham}, {Bland-Hawthorn}, {Jones}, {Sharp},
  {Thomas}, {Liske}, {Loveday}, {Norberg}, {Peacock}, {Bamford}, {Brough},
  {Colless}, {Cameron}, {Conselice}, {Croom}, {Frenk}, {Gunawardhana},
  {Kuijken}, {Nichol}, {Parkinson}, {Phillipps}, {Pimbblet}, {Popescu},
  {Prescott}, {Sutherland}, {Tuffs}, {van Kampen}, \&
  {Wijesinghe}}]{Taylor2011}
{Taylor}, E.~N., {Hopkins}, A.~M., {Baldry}, I.~K., {et~al.} 2011, \mnras, 418,
  1587

\bibitem[{{Tinker}(2021)}]{Tinker2021}
{Tinker}, J.~L. 2021, \apj, 923, 154

\bibitem[{{Tinker}(2022)}]{Tinker2022}
{Tinker}, J.~L. 2022, \aj, 163, 126

\bibitem[{{Truong} {et~al.}(2023){Truong}, {Pillepich}, {Nelson}, {Bogd{\'a}n},
  {Schellenberger}, {Chakraborty}, {Forman}, {Kraft}, {Markevitch},
  {Ogorzalek}, {Oppenheimer}, {Sarkar}, {Veilleux}, {Vogelsberger}, {Wang},
  {Werner}, {Zhuravleva}, \& {Zuhone}}]{Truong2023}
{Truong}, N., {Pillepich}, A., {Nelson}, D., {et~al.} 2023, \mnras, 525, 1976

\bibitem[{{Truong} {et~al.}(2021){Truong}, {Pillepich}, {Nelson}, {Werner}, \&
  {Hernquist}}]{Truong2021b}
{Truong}, N., {Pillepich}, A., {Nelson}, D., {Werner}, N., \& {Hernquist}, L.
  2021, \mnras, 508, 1563

\bibitem[{{Tumlinson} {et~al.}(2011){Tumlinson}, {Thom}, {Werk}, {Prochaska},
  {Tripp}, {Weinberg}, {Peeples}, {O'Meara}, {Oppenheimer}, {Meiring}, {Katz},
  {Dav{\'e}}, {Ford}, \& {Sembach}}]{Tumlinson2011}
{Tumlinson}, J., {Thom}, C., {Werk}, J.~K., {et~al.} 2011, Science, 334, 948

\bibitem[{{Velander} {et~al.}(2014){Velander}, {van Uitert}, {Hoekstra},
  {Coupon}, {Erben}, {Heymans}, {Hildebrandt}, {Kitching}, {Mellier}, {Miller},
  \& et~al.}]{VelandervanUitertHoekstra_2014MNRAS.437.2111V}
{Velander}, M., {van Uitert}, E., {Hoekstra}, H., {et~al.} 2014, \mnras, 437,
  2111

\bibitem[{{Vladutescu-Zopp} {et~al.}(2023){Vladutescu-Zopp}, {Biffi}, \&
  {Dolag}}]{Vladutescu-Zopp2023AA...669A..34V}
{Vladutescu-Zopp}, S., {Biffi}, V., \& {Dolag}, K. 2023, \aap, 669, A34

\bibitem[{{Wang} {et~al.}(2016){Wang}, {Li}, {Jiang}, \& {Fang}}]{Wang2016}
{Wang}, Q.~D., {Li}, J., {Jiang}, X., \& {Fang}, T. 2016, \mnras, 457, 1385

\bibitem[{{Whitaker} {et~al.}(2012){Whitaker}, {van Dokkum}, {Brammer}, \&
  {Franx}}]{Whitaker2012}
{Whitaker}, K.~E., {van Dokkum}, P.~G., {Brammer}, G., \& {Franx}, M. 2012,
  \apjl, 754, L29

\bibitem[{{White} \& {Rees}(1978)}]{White1978}
{White}, S.~D.~M. \& {Rees}, M.~J. 1978, \mnras, 183, 341

\bibitem[{{Wright} {et~al.}(2024){Wright}, {Somerville}, {Lagos}, {Schaller},
  {Dav{\'e}}, {Angl{\'e}s-Alc{\'a}zar}, \& {Genel}}]{Wright2024}
{Wright}, R.~J., {Somerville}, R.~S., {Lagos}, C. d.~P., {et~al.} 2024, arXiv
  e-prints, arXiv:2402.08408

\bibitem[{{Zarrouk} {et~al.}(2022){Zarrouk}, {Ruiz-Macias}, {Cole}, {Norberg},
  {Baugh}, {Brooks}, {Gazta{\~n}aga}, {Kitanidis}, {Kehoe}, {Landriau},
  {Moustakas}, {Prada}, \& {Tarl{\'e}}}]{Zarrouk2022}
{Zarrouk}, P., {Ruiz-Macias}, O., {Cole}, S., {et~al.} 2022, \mnras, 509, 1478

\bibitem[{{Zehavi} {et~al.}(2011){Zehavi}, {Zheng}, {Weinberg}, {Blanton},
  {Bahcall}, {Berlind}, {Brinkmann}, {Frieman}, {Gunn}, {Lupton}, \&
  et~al.}]{ZehaviZhengWeinberg_2011ApJ...736...59Z}
{Zehavi}, I., {Zheng}, Z., {Weinberg}, D.~H., {et~al.} 2011, \apj, 736, 59

\bibitem[{{Zhang} {et~al.}(2024){Zhang}, {Comparat}, {Ponti}, {Meloni},
  {Nandra}, {Haberl}, {Locatelli}, {Zhang}, {Sanders}, {Zheng}, {Liu},
  {Popesso}, {Liu}, {Truong}, {Pillepich}, {Predehl}, \&
  {Salvato}}]{Zhang2024profile}
{Zhang}, Y., {Comparat}, J., {Ponti}, G., {et~al.} 2024, arXiv e-prints,
  arXiv:2401.17308

\bibitem[{{Zou} {et~al.}(2019){Zou}, {Gao}, {Zhou}, \& {Kong}}]{Zou2019}
{Zou}, H., {Gao}, J., {Zhou}, X., \& {Kong}, X. 2019, \apjs, 242, 8

\end{thebibliography}

\end{document}